\documentclass{aa}
\usepackage{graphicx}
\usepackage{natbib}
\begin{document}
%

   \title{Resolved Hubble Space spectroscopy of ultracool binary systems 
\thanks{Based on observations obtained with the NASA/ESA 
\emph{Hubble Space Telescope}}}


   \author{E.~L. Mart\'\i n \inst{1,2}, 
W. Brandner \inst{3}, 
H. Bouy \inst{1}, 
G. Basri \inst{4}, 
J. Davis \inst{2}, R. Deshpande\inst{2}, M. Montgomery \inst{2}  
}

   \offprints{E.~L. Mart\'\i n}

   \institute{Instituto de Astrof\'\i sica de Canarias, 38200 La Laguna, Spain\\
              \email{ege@iac.es}
   \and University of Central Florida, Dept. of Physics, PO Box 162385, 
         Orlando, FL 32816-2385, USA 
              \and Max-Planck Institut f\"ur Astronomie, K\"onigstuhl 17, 
D-69117 Heidelberg, Germany
\and Astronomy Department, University of California, Berkeley, CA 94720, USA \\ 
}

  \date{}

   \abstract{
Using the low-resolution mode of the Space Telescope Imaging 
Spectrograph (STIS) 
aboard the  \emph{Hubble Space Telescope} (HST), 
we have obtained spatially resolved spectra of 20 ultracool dwarfs. 
18 of them belong to 9 known very low-mass binary systems with 
angular separations in the range 0\farcs37-0\farcs098. 
We have derived spectral types in the range dM7.5 to dL6 
from the PC3 index, and by comparing our STIS spectra with 
ground-based spectra of similar spectral resolution 
from Mart\'\i n et al. (1999). 
We have searched for H$_\alpha$ emission in each 
object but it was clearly detected in only 2 of them. We find that the 
distribution of  H$_\alpha$ emission in our sample is statistically 
different from that of single field dwarfs, suggesting an intriguing  
anticorrelation between chromospheric activity and binarity for M7--M9.5 dwarfs.     
We provide measuments of the strength of the main photospheric features and 
the PC3 index, and we derive calibrations of spectral subclasses versus 
F814W and K-band absolute magnitudes 
for a subset of 10 dwarfs in 5 binaries that have 
known trigonometric parallaxes.   
 
}

\authorrunning{Mart\'\i n et al.}
\titlerunning{Resolved HST spectroscopy of ultracool binaries}

\maketitle

\markboth{STIS observations of ultracool binaries}{Mart\'\i n et al.}

\section{Introduction}

Late-type nearby spatially resolved binaries 
(usually referred to as visual binaries) have been 
patiently followed up for many decades to obtain orbital parameters 
(e.g., van de Kamp 1938; Couteau 1957; Chang 1972; Abt \& Levy 1973). 
The dynamical masses obtained from the studies of 
resolved binaries have been used to calibrate the mass-luminosity relationship 
for low-mass stars (e.g., Delfosse et al. 2000). 
A number of recent surveys have extended the search for resolved binaries to 
very low-mass (VLM) stars and brown dwarfs (BDs). 
Hubble Space Telescope imaging has been used to search for 
VLM binaries\footnote{In this paper 
we adopt the definition that a VLM binary has a primary with spectral 
type M6 and later.} 
in young associations and open clusters 
(Mart\'\i n et al. 2000b, 2003; Kraus et al. 2005) and in the solar 
neighborhood 
(Mart\'\i n et al. 1998, 1999; Reid et al. 2001; Bouy et al. 2003; 
Burgasser et al. 2003; Gizis et al. 2003; Golimowski et al. 2004). 
Near infrared-imaging on large ground-based telescopes, usually assisted 
by adaptive optics,  
has also been effective in finding VLM binaries 
(e.g., Koerner et al. 1999; Mart\'\i n et al. 2000a; Close et al. 2003; 
Potter et al. 2002; Siegler et al. 2005). 
These studies have determined that the binary frequency among field 
VLM stars and BDs in 
the separation range 1-15 AU is about 15\% , and that there is 
a sharp drop in the binary frequency for separations larger than 15 AU. 
Nevertheless, a few examples of wider field ultracool binaries 
are known (Mart\'\i n et al. 1998, 2000b; 
Bill\`eres et al. 2005; Phan Bao et al. 2005; Burgasser \& McElwain 2006).  
On the other hand, 
it is not clear what the VLM binary frequency is within 1 AU because 
of the lack of long-term spectroscopic binary surveys for VLM stars 
and BDs. 
The statistical properties of VLM binaries will continue to be investigated 
because 
they provide a fundamental constraint for models 
of VLM star formation (Kroupa et al. 2003; Umbreit et al. 2005). 

The increasing number of VLM binaries can be used for follow-up studies of 
their properties. 
Astrometric monitoring of three VLM binaries have yielded the first 
estimates of the 
orbital parameters and dynamical masses (Lane et al. 2001; 
Bouy et al. 2004; Brandner et al. 2004). 
Spatially-resolved low-resolution spectroscopic follow-up has been made in 
a few cases 
and spectral types have been obtained (Lane et al. 2001; Goto et al. 2002; 
Potter et al. 2002; Bouy et al. 2004; Chauvin et al. 2004; Luhman 2004, 2005; 
McCaughrean et al. 2004; Billeres et al. 2005; and Burgasser \& McElwain 2006). 
Resolved high-resolution spectroscopic observations have been reported for 
only one VLM binary, namely GJ~569~B, 
and have allowed to dynamically estimate the individual masses of each component 
(Zapatero Osorio et al. 2004; Simon et al. 2006). Accurate dynamical masses 
for an eclipsing BD binary have also been reported recently (Stassun et al. 2006).
   
In this paper, 
we present resolved low resolution spectroscopic observations of 
9 VLM binaries, 
which represent a significant increase in the number of members of VLM systems 
that have been 
characterized spectroscopically.  In section 2, we describe the selection of 
the sample, the STIS observations and the processing of the data. 
In section 3 we describe the spectra, and we present the determination of the 
spectral types. Section 4 deals with the discussion of 
 chromospheric activity and the main absorption features in our sample, 
and we derive calibrations of spectral subclass versus absolute 
magnitudes for the components of VLM binaries with known parallaxes. 
Finally, section 5 summarizes our results.

\section{Observations and data processing \label{obs}}

Our sample was selected from known VLM binaries with separations in 
the range 0\farcs37-0\farcs098 and I-band magnitude brighter than I=19, 
so that resolved optical low-resolution spectroscopy of 
each component could be obtained with STIS.  
Table~1 provides the STIS observing log. 
The observations of 2MASSW~J07464256+2000321 
have already been presented in Bouy et al. (2004) 
but we also include them in our analysis for completeness.  

Our goal was to obtain spatially resolved spectra of each component of 
the binary systems. 
However, scheduling constraints of HST made it impossible to 
perfectly align the slit along the axis of each binary. 
In one case, namely DENIS-P~J144137.3-094559, 
we could not get any spectrum of the secondary because the slit 
was not oriented along the binary semimajor axis. 
Another case, namely 2MASSW~J0920122+351742, 
turned out not to be resolved at the epoch of observations. 
It is possible that this object may not be a binary because it has been 
resolved in only one epoch 
(Bouy et al. 2006, in preparation). For the other binaries we obtained 
resolved 2-D spectra and performed the extraction of the two spectra as 
explained in detail in Bouy et al. (2004). 
An example of the de-blending of spectra in two of our binary systems 
(including one of the tightest) is shown in Figure 1. 
We used the bias, flat field and wavelength calibrations provided by the 
HST pipeline. 

The grating used was the G750L centered at 775.1 nm, and the aperture 
was 52\farcs0 ~long by 0\farcs2 
~wide.  The total wavelength range of each spectrum is 525--1100~nm, 
but the usable range depends 
on the brightness and the spectral type of each object. 
Residual fringing is clearly present at wavelengths longer than about 900~nm.  
The nominal dispersion is 4.92 \AA/pixel and according to the STIS handbook, 
our slit width is expected to project onto 4 pixels (FWHM = 20 \AA ).  
For comparison the resolution quoted by 
Kirkpatrick et al. (1999)  was 9 \AA , and the resolution of the spectra presented in  
Mart\'\i n et al. (1999) (hereafter M99) was 12 \AA . 
Thus, our STIS data have a resolution somewhat lower than those papers.  
In Figure 2 we display final spectra for three of our targets spanning 
the whole range of spectral subclasses in our sample.

\section{Spectral classification}

Our resolved STIS spectra of VLM binaries allowed us to determine the 
spectral type of each component. Only two 
of our program binaries have previous determinations of spectral type 
for each component:  
Lane et al. (2001) gave spectral types of M8.5 and M9 for GJ~569~Ba and Bb, 
respectively, using low-resolution near-infrared ($J$-band) spectra. 
Bouy et al. (2004) obtained spectral types of dL0 and dL1.5 for 
2MASSW~J0746425+200032~A and B, 
respectively, using the same STIS data as this work. 
They compared the STIS spectra 
with ground-based low-resolution spectra from  M99. 

In this work we have estimated the spectral subclasses using two methods: 
(a) we measured the PC3 index defined by M99 and used their PC3-spectral 
type relationship for M and L dwarfs, and (b) we compared our spectra with 
those of M99 and chose the best match. The spectral subclasses obtained 
from both approaches are given in Table 2. When the subclasses from the 
two methods agreed with 1 subclass, we computed the average and we 
rounded in steps of 0.5 subclass. When there was a disagreement, we 
favored the subclass derived from method (b). This happened only for the 
latest L dwarfs of our sample.

\subsection{Description of the spectra of each binary}

In this subsection the STIS spectra of each target 
are displayed and they are compared with the best matching M99 spectra:    

\begin{itemize} 

\item DENIS-P~J020529.0-115925~A (dL5): 
The K~\textsc{i} resonance doublet is slightly broader, and 
the CrH and FeH bands are weaker than in the M99 spectrum 
of DENIS-P~J1228-2415 (Figure 3). The M99 spectrum of DENIS-P~J0205-1159 
is the best match to our STIS spectrum of DENIS-P~J020529.0-115925~A, indicating, 
as expected, that the primary dominates the composite ground-based spectrum.  
We note that the PC3 index gives a slightly earlier subclass of dL4.1 
than the spectrum matching technique. 

\item DENIS-P~J020529.0-115925~B (dL6): The CrH and FeH bands have almost vanished. 
The M99 spectrum of DENIS-P~J0255-4700 is the best match except for 
the steam band at 930~nm, which is stronger in the ground-based spectrum (Figure 3). 
This could be due to the contribution of telluric absorption in the M99 spectrum.      
We also note that the STIS spectra are noisy in this region due to low quantum 
efficiency of the detector and the presence of  
a strong interference pattern at these red wavelengths. 
The PC3 index gives a spectral type 2 subclasses earlier 
than the spectrum match, suggesting 
that this index is not reliable for the latest L dwarfs. 

\item DENIS-P~J035726.9-441730~A (dM9): As shown in Figure~4, the best match 
spectrum from M99 is that of DENIS-P~1431-1953 (dM9). In the ground-based spectrum 
the region from 740~nm to 770~nm is affected by telluric absorption. 
The PC3 index also gives a subclass of dM9 in perfect agreement with 
the spectrum match. 

\item DENIS-P~J035726.9-441730~B (dL1.5): As shown in Figure~4, the best match 
spectrum from M99 is that of DENIS-P~1441-0945 (dL1). In the ground-based spectrum 
the region from 740~nm to 770~nm is affected by telluric absorption so it is not 
surprising the mismatch between the STIS data and the M99 data over this wavelength 
coverage. A similar discrepancy is seen in the spectra of all our late-M and early-L 
dwarfs.  
The PC3 index gives a slightly later subclass of dL1.6. We averaged the two 
methods and rounded our adopted subclass to dL1.5.  

\item DENIS-P~J100428.3-114648~A (dM9.5): Both the PC3 index and the best match 
spectrum (DENIS-P~1208+0149) give a spectral subclass of dM9.5 (Figure 5). 

\item DENIS-P~J100428.3-114648~B (dL0.5): The PC3 index gives dM9.9 but the 
STIS spectrum appears to intermediate between a dL0 (DENIS-P~0909-0658)  
and a dL1 (DENIS-P~1441-0945) as illustrated in Figure~5. We adopt a subclass 
of dL0.5 by interpolating between dL0 and dL1. 

\item DENIS-P~J122821.6-241541~A (dL4): The PC3 index gives dL3.9 and 
the M99 spectrum of LHS~102~B (dL4) is the best spectral match (Figure 6). 
The Li~\textsc{i} resonance doublet is not detected 
even though it has been clearly seen from the ground 
(Mart\'\i n et al. 1997; Tinney et al. 1997). 
This non detection is due to the low signal to noise ratio and low resolution 
of our STIS data. 

\item DENIS-P~J122821.6-241541~B (dL4.5): The PC3 index gives dL4.2 and 
the M99 spectrum of DENIS-P~J1228-2415 (bdL4.5) is the best spectral match 
(Figure 6).    
The Li~\textsc{i} resonance doublet is not detected because of the 
low signal to noise ratio of our data at 670.8~nm.  

\item DENIS-P~J144137.3-094559~A (dL1):  
The PC3 index gives dL1.3, and the best M99 match is DENIS-P~J1441-0945 itself 
(Figure 7). We note that    
DENIS-P~J144137.3-094559 is a common proper motion companion of the star  
G124-62  (dM4.5, Seifahrt, Guenther \& Neuhauser 2005), which 
is a member of the Hyades supercluster. 

\item GJ~569~Ba (dM9): The PC3 index gives a subtype of dM9.2 and the 
best spectral match is DENIS-P~1431-1953 (dM9) so there is  good agreement 
between the two methods.   

\item GJ~569~Bb (dM9):  The PC3 index gives a subtype of dM8.6, but 
the best match is also DENIS-P~1431-1953 (Figure 8). 
This example underlines the benefits of using several criteria to determine accurate 
spectral subclasses among  ultracool dwarfs. 
If the spectral class were assigned using only the PC3 index, 
we would give a slightly earlier subclass to GJ~569~Bb than to GJ~569~Ba, 
which is not consistent with the properties of this binary. 
 
\item 2MASSW~J0746425+200032~A (dL0): 
The PC3 index gives a subclass of dM9.5 but the best match is DENIS-P~0909-0658 
(dL0) as shown in Bouy et al. (2004).  
The CrH$\lambda$861.1~nm and FeH$\lambda$869.2~nm bands are stronger than in M dwarfs.  
However, the CrH and FeH bands are not stronger than the nearby TiO band at 843.2~nm. 

\item 2MASSW~J0746425+200032~B (dL1.5): 
The PC3 index gives a subclass of dM9.6 but the best match is 
intermediate between DENIS-P~1441-0945  
(dL1) and Kelu~1 (dL2), as shown in Bouy et al. (2004).    
The TiO and VO bands are weaker, and the . 
CrH and FeH bands are stronger than in the A component. 

\item 2MASSW~J0920122+351742 (dL5): 
The PC3 index gives a subclass of dL5.2 and the best match is DENIS-P~J0205-1159 
(dL5) as shown in Figure~9, so there is good agreement between the two methods.

\item 2MASSW~J1146344+223052~A (dL2): 
The PC3 index gives a subclass of dL1.4 and the best spectral match is Kelu~1 
(bdL2) as shown in Figure~10.  
The K~\textsc{i} resonance lines become so broad that the doublet is blended. 
They are clearly broader than for the dL1.  
The CrH and FeH bands are stronger than for the dL1 and the TiO bands are weaker. 

\item 2MASSW~J1146344+223052~B (dL2): 
The PC3 index gives a subclass of dL1.7 and the best spectral match is Kelu~1 
(bdL2) as shown in Figure~10. Bouy et al. (2003) reported a magnitude difference 
in the F814W filter of 0.75 mag., which is confirmed by new ACS images 
(H. Bouy 2006, private communication).   
The similar spectral types but different brightness 
of these two dwarfs suggests that 2MASSW~J1146344+223052~A could be 
an unresolved binary with 
nearly equal components. Thus, 2MASSW~J1146344+223052 may be a triple system. 
Very low-mass triple systems may not be rare as indicated by the recent observations of 
GJ~900 (Mart\'\i n 2003) and DENIS-P J020529.0-115925 (Bouy et al. 2005). 
Confirmation of the suspected 
unresolved binarity of 2MASSW~J1146344+223052~A requires high angular 
resolution observations or radial velocity monitoring. 

\item 2MASSW~J1311391+803222~A,B (dM7.5, dM8). As shown in Figure 11, 
the STIS spectra of both components are nearly identical. The 
spectral type inferred from the PC3 index are dM7.3 for A and 
dM7.6 for B. The best spectral match for the A component is LHS2243 (dM7.5)  
and for the B component is VB10 (dM8). 
The two lines of the K~\textsc{i} resonance doublet can be distinguished 
with our spectral resolution. 
The 766.5~nm line is stronger than the 769.9~nm line. 
The Cs~\textsc{i} is too weak to be seen in our spectra. 
The TiO and VO bands are conspicuous and the small 0.5 subclass difference 
is best seen in the VO band at 740~nm.

\item 2MASSW~J1426316+155701~A (dM8) and B (dL1.5):   
The spectrum of component A is very similar to that of VB10, which 
agrees with the dM8 subclass obtained from the PC3 measurement, while 
the spectrum of the component B is intermediate between that of 
DENIS-P~J1441-0945 (dL1) and Kelu~1 (bdL2) as illustrated in  
Figure~12.

\end{itemize} 

\section{Discussion}

\subsection{Chromospheric activity and binarity}

H$_\alpha$ emission is an indicator of chromospheric activity due to 
nonthermal heating by magnetic fields. 
We have detected H$_\alpha$ emission in only 2 of our objects. We 
give the equivalent widths measurements or upper limits in Table~3. 
We did not see any flares, which usually display a variety of strong 
emission lines in late-M and L dwarfs (Liebert et al. 1999; Mart\'\i n 1999; 
Mart\'\i n \& Ardila 2001). 

In Figure~13, we display a zoom of the H$_\alpha$ region for most of our targets. 
Our three coolest L dwarfs (dL5--dL6) are missing from the plot because they do not have 
enough continuum for equivalent width determination at these wavelengths. 
We have compared the distribution of H$_\alpha$ emission equivalent width 
with respect to spectral subclass in our sample 
with the measurements obtained by Gizis et al. (2000) for field dwarfs. 
We performed a Kolmogorov-Smirnov test to check whether the two distributions 
of H$_\alpha$ emission equivalent width values are statistically 
different or similar. We binned the data in steps of 0.5 subclasses, so there 
were 6 spectral bins in the range dM7--dM9.5. We found that  
the two distributions of  
spectral types are 90\% similar, but their associated distributions of 
H$_\alpha$ equivalent widths are different with high level of confidence (99.9\%). 
Our sample of resolved binary components tends to show lower H$_\alpha$ 
activity than single field dwarfs in the spectral range dM7--dM9.5. 
The situation is unclear for L dwarfs because there are too few datapoints.  

Our results indicate that there could 
be an anticorrelation between chromospheric activity and resolved binarity for the 
latest dMs. 
This connection may be due to different angular momentum histories in the binary 
components, although this is not very likely because there is not a good 
connection between activity and rotation for these late dwarfs (Mohanty \& Basri 2003). 
On the other hand, chromospheric activity has been suggested to be enhanced 
in some late-type dwarfs after impact of asteroids or comets 
(e.g. AB Dor, G\'omez de Castro 2002). A ``planetesimal-impact'' hypothesis to 
explain flares in cool stars has also been discussed in the literature 
(Hertzsprung 1924; Andrews 1991). 
Our results may lend support to this 
hypothesis because in the binaries of our sample there may be less debris 
material due to disk clearing by the components of the binary system. 
In single very-late dMs, debris material may last over  periods of time 
longer than for solar-type stars because of the reduced effect of disk dissipation 
processes such as the Poynting-Robertson drag.

\subsection{Alkali lines}

We searched for the 
Li~\textsc{i} resonance line in our spectra. 
This line is a useful diagnostic of the age and mass of ultracool dwarfs 
(Magazz\`u et al. 1993; Mart\'\i n et al. 1994) 
and it has been detected in the unresolved spectra of 
one of our program binaries with equivalent widths in the range 2-4 \AA ~
(DENIS-P~J122821.6-241541~A,B; Mart\'\i n et al. 1997; 
Tinney, Delfosse \& Forveille 1997). 
However, we could not detect it in our spectrum of DENIS-P~J122821.6-241541~A. 
We conclude that the resolution and limited signal to noise ratio of 
our STIS spectra do not allow us to detect Li~\textsc{i} resonance lines with equivalent 
widths smaller than about 5 \AA , 
implying that we cannot set useful constrains of the lithium abundance of our 
program dwarfs.    

We measured the K~\textsc{i} resonance doublet at 766.5 and 769.9~nm. 
For the dwarfs where we could distinguish the two lines, we used the gaussian 
deblending option in the IRAF package splot. 
The equivalent widths obtained are given in Table 3 and their dependence with 
spectral class is illustrated in Figure~14. 
At subclass dL2 and later the doublet becomes extremely broad and the 
two lines are blended 
(Mart\'\i n et al. 1997, 1999; Burrows \& Sharp 1999; Kirkpatrick et al. 1999). 
For those dLs we provide the equivalent width of the blended pair. 
The Na~\textsc{i} subordinate doublet at 818.4 and 819.5~nm could not be 
measured because of a 
bad column in the array. The Cs~\textsc{i} at 852.1 and 894.3~nm were strong 
enough in the 
coolest objects to measure the equivalent width (Table 3).  

The scatter in the equivalent widths of the K~\textsc{i} resonance doublet as a function 
of spectral class that is seen in Figure~14 is likely 
       due to its well-known sensitivity to surface gravity (M99). 
For example, we note that GJ~569~Ba and Bb (dM9) have weaker K~\textsc{i} equivalent width 
than 2MASSW~J1426316+155701~A (dM8) and DENIS-P~J100428.3-114648~A (dM9.5). Weaker 
K~\textsc{i} is an indication of lower surface gravity, and 
consequently a younger age and lower mass for a given spectral subclass. 
The age of GJ~569~B has been estimated to be young ($\sim$300 Myr) 
by Zapatero Osorio et al. (2004) using evolutionary tracks. It is likely that the ages 
of DENIS-P~J100428.3-114648 and 2MASSW~J1426316+155701 are older than that of GJ~569~B. 
With more observations and detailed modelling, 
 the K~\textsc{i} resonance doublet may yield a useful age 
calibration for very low-mass stars and brown dwarfs.

\subsection{Molecular bands}

The main molecular bands in our spectra are labeled in Figure 2. 
We measured their strengths using the indices defined in 
M99, and we give the values in Table 4. 
As already discussed in Kirkpatrick et al. 1999 and 
M99, the TiO bands diminish in strength from the late-M to the L dwarfs, 
and eventually become undetectable in mid-L dwarfs. 
Their dependence on spectral type is shown in  Figure 15.                   .
This effect is understood in terms of the settling of Ti onto dust grains 
such as CaTiO$_{\rm3}$, 
which condenses at temperatures below 2500~K (Allard et al. 2001).  
The VO molecule behaves in a similar way to the TiO but it disappears at 
slightly cooler temperature (Figure 15). 
The chromium and iron hydrides become prominent in mid-L dwarfs, 
and tend to fade away in the late-L dwarfs (Figure 15). 

The values of the two TiO indices are lower in GJ~569~Ba than 
in our dM7.5-dM8 targets, but the VO indices are more similar.  
This may be another spectroscopic manifestation of low surface gravity. 
The weakening of TiO bands may be shifted to later spectral subclass in 
low gravity objects because of less efficient dust formation. 
It has been noted that dM7-dM9 members of the young Pleiades cluster have 
stronger TiO bands 
than their older counterparts in the field (Mart\'\i n et al. 1996).


\subsection{Comparison of spectral subclasses}

In Table~5 we show a comparison of the spectral subclasses adopted by us in the M99 system 
with those reported in the literature. The spectral types given in the literature correspond 
to the unresolved systems, and thus are dominated by the primary. There is a fairly good 
agreement. No discrepancies larger than 2 subclasses are noted. 
The near-infrared spectral types 
from Geballe et al. (2002) for two of our binaries also agree within 2 subclasses. 
The near-infrared spectral types of Gl~569~Ba and b reported by Lane et al. (2001) 
are consistent with our optical spectral types within the uncertainties (half 
a subclass). 

\subsection{Absolute magnitude versus spectral type relationships}

Among our sample, only 5 binaries have known parallaxes. Their properties are summarized 
in Table~6. In Figure 16 we plot the spectral types of their resolved components versus  
the absolute magnitudes in the photometric bands $F814W$ and $K_s$.  As expected later 
type objects are cooler and have fainter magnitudes. Using second order polynomials, 
we find that the following equations provide a good fit to our data in the 
spectral class range dM9 -- dL6: 

\begin{equation} 
{M_{F814W} =  - 2.2167  + 2.3284 \times SpT  - 0.0682 \times SpT^2  }
\end{equation} 


\begin{equation}
{M_{Ks} = 10.502 - 0.2311 \times SpT + 0.0226 \times SpT^2 }
\end{equation} 
 
where we used the numerical code of SpT=9 for M9 through SpT=16 for L6. 
These fits are shown as dotted lines in Figure~16. 
The scatter (1~$\sigma$) in relations 1 and 2 over the sample of objects that we fit
these relations to is 0.43 mag. and 0.27 mag., respectively. 
The good correlation between our adopted spectral types and the 
absolute magnitudes indicates 
that the late-M and L dwarf spectral classification scale depends primarily on effective temperature. 
The observed scatter around these relations 
is likely due to the added contributions of unresolved binarity (higher order multiple systems), 
observational errors, gravity effects and metallicity differences in the sample. 
Our results are consistent with those reported by other authors such 
as Kirkpatrick et al. (2000), Dahn et al. (2002) and Vrba et al. (2004) 
within the observational error bars. 
A comparison of our SpT -- M$_{K_s}$ 
relationship with that of Vrba et al. (2004) is shown in Figure~16.

\section{Summary}

We have presented low resolution (R=470) optical spectroscopy of 20 ultracool dwarfs in resolved 
binary systems. 18 targets are members of 9 resolved binaries with 
angular separations in the range 0\farcs37-0\farcs098. 

We derived spectral types for our targets using the PC3 index and direct comparison 
with the M99 ground-based spectra. 
We report that the H$_\alpha$ emission in our targets is statistically weaker 
than that of     field dwarfs for the range of spectral class dM7--dM9.5. 
We did not detect any flare. 
We did not detect the Li~\textsc{i} resonance doublet in our targets because of the poor 
quality and low resolution of our STIS spectra. We provide pseudo-equivalent widths of the 
Cs~\textsc{i} and K~\textsc{i} resonance doublets for the targets where those lines 
could be measured. These lines tend to increase in breath and 
strength toward later spectral types, as already reported by Kirkpatrick et al. (1999) and 
M99, but there is significant dispersion which may be due to gravity 
effects. This doublet 
may serve as a useful age indicator for field ultracool dwarfs. 

For a subset of 10 targets in 5 binaries with known trigonometric parallaxes,  
we show that there is a good correlation between our spectral types and the absolute magnitudes 
of the targets in the $F814W$ and $K_s$ bands. We provide second order polynomial fits 
to the data, which could be used to derive spectrophotometric parallaxes of late-M and L 
field dwarfs. 

After publication of this paper, we plan to make the spectra available online via the IAC 
catalog of ultracool dwarfs. The address of this catalog is: 
{\bf http://www.iac.es/galeria/ege/catalogo$_{-}$espectral/}. The description 
of the catalog can be found in  
Mart\'\i n, Cabrera \& Cenizo (2005).

\begin{acknowledgements}
We thank Jay Anderson and the staff at STScI for help with the STIS observations.       
This work is based on observations collected  with the NASA/ESA Hubble Space 
Telescope operated at the Space 
Telescope Science Institute (STSci), programs GO9157, GO9451, and GO9499.  
The STSci is operated by the Association of Universities for 
Research in Astronomy, Inc., under NASA contract NAS 5-26555. 
Research presented herein was partially funded by NSF research grant AST 02-05862, 
by the Deutsches Zentrum f\"ur Luft- und Raumfahrt (DLR),
F\"orderkennzeichen 50 OR 0401, and by Spanish MEC AYA2005-06453. 
\end{acknowledgements}

\bibliographystyle{aa}


\clearpage

\renewcommand{\arraystretch}{1.5}


  \begin{table}
	      \caption[]{STIS observing log.} 
         \label{obs_log}
     \begin{tabular}{lccccc}
            \hline
	    \hline

            Object               &  Exp.      & Date Obs.           & Program   & Ref. Object & Ref. Binarity \\
	                         & Time [s]   & \textsc{dd/mm/yyyy} &           & Discovery  & Discovery\\
\hline
DENIS-P~J144137.3-094559~A        & 4695       & 29/03/2002          & GO9157   & Mart\'\i n et al. (1999) & 
Bouy et al. (2003) \\
DENIS-P~J122821.6-241541~A,B      & 4693       & 25/04/2002          & GO9157   & Delfosse et al. (1998) & Mart\'\i n et al. (1999b) \\
GJ~569~Ba,Bb                      & 3873       & 26/06/2002          & GO9499    & Forrest et al. (1988) & Mart\'\i n et al. (2000a) \\ 
DENIS-P~J020529.0-115925~A,B      & 4695       & 25/09/2002          & GO9157   & Delfosse et al. (1997) & Koerner et al. (1999) \\
DENIS-P~J035726.9-441730~A,B      & 4895       & 08/01/2003          & GO9451   & Bouy et al (2003) & 
Bouy et al. (2003) \\
2MASSW~J1146344+223052~A,B        & 4702       & 10/02/2003          & GO9157   & Kirkpatrick et al. (1999) & Koerner et al. (1999) \\
2MASSW~J1311391+803222~A,B        & 4183       & 27/02/2003          & GO9451   & Gizis et al. (2000) & Close 
et al. (2003) \\
2MASSW~J0920122+351742            & 4774       & 10/03/2003          & GO9451    & Kirkpatrick et al. (2000) & 
Bouy et al. (2003) \\ 
DENIS-P~J100428.3-114648~A,B      & 4324       & 14/03/2003          & GO9451   & Bouy et al (2003)  & 
Bouy et al. (2003) \\
2MASSW~J1426316+155701~A,B        & 1980       & 28/04/2003          & GO9451   & Gizis et al. (2000) &  
Close et al. (2003)  \\
2MASSW~J0746425+200032~A,B        & 1980       & 09/01/2004          & GO9451   & Kirkpatrick et al. (2000) & 
Reid et al. (2001) \\
            \hline
     \end{tabular}
   \end{table}   


\clearpage

\begin{table*}
\caption[]{PC3 index and spectral subclasses}
\label{atomic_lines}
\begin{tabular}{ lcccc }
\hline
\hline
Name of                     & PC3   & SpT   &  SpT          & SpT          \\
\hline
Object                      &       & (PC3) &  (best match) & (adopted)     \\
\hline
DENIS-P~J020529.0-115925~A  & 6.21  & dL4.1 & dL5           & dL5           \\ 
DENIS-P~J020529.0-115925~B  & 5.48  & dL3.6 & dL6           & dL6           \\
DENIS-P~J035726.9-441730~A  & 2.15  & dM9.1 & dM9           & dM9           \\ 
DENIS-P~J035726.9-441730~B  & 2.94  & dL1.6 & dL1           & dL1.5         \\
DENIS-P~J100428.3-114648~A  & 2.33  & dM9.6 & dM9--dM9.5    & dM9.5         \\
DENIS-P~J100428.3-114648~B  & 2.49  & dM9.9 & dL0--dL1      & dL0.5         \\
DENIS-P~J122821.6-241541~A  & 5.92  & dL3.9 & dL4           & dL4           \\
DENIS-P~J122821.6-241541~B  & 6.39  & dL4.2 & dL4.5         & dL4.5         \\ 
DENIS-P~J144137.3-094559~A  & 2.56  & dL1.3 & dL1           & dL1           \\   
GJ~569~Ba                   & 2.17  & dM9.2 & dM9           & dM9           \\
GJ~569~Bb                   & 1.99  & dM8.6 & dM9           & dM9           \\
2MASSW~J0746425+200032~A    & 2.29  & dM9.5 & dL0           & dL0           \\   
2MASSW~J0746425+200032~B    & 2.34  & dM9.6 & dL1.5         & dL1.5         \\   
2MASSW~J0920122+351742      & 8.58  & dL5.2 & dL5           & dL5           \\  
2MASSW~J1146344+223052~A    & 2.69  & dL1.4 & dL2           & dL2           \\   
2MASSW~J1146344+223052~B    & 3.07  & dL1.7 & dL2           & dL2           \\  
2MASSW~J1311391+803222~A    & 1.68  & dM7.3 & dM7.5         & dM7.5         \\  
2MASSW~J1311391+803222~B    & 1.75  & dM7.6 & dM8           & dM8           \\
2MASSW~J1426316+155701~A    & 1.78  & dM7.7 & dM8           & dM8           \\ 
2MASSW~J1426316+155701~B    & 2.75  & dL1.4 & dL1--dL2      & dL1.5         \\ 
\hline
\end{tabular}
\end{table*}

\clearpage

\begin{table*}
\caption[]{Atomic line data}
\label{atomic_lines}
\begin{tabular}{ lccccc }
\hline
\hline
Name of                    &   H$_\alpha$  &  K~\textsc{i}  & K~\textsc{i}& Cs~\textsc{i}  & Cs~\textsc{i} \\
\hline
Object                     & 656.3                &   766.5               &   769.9  &  852.1   &   894.3   \\
\hline
DENIS-P~J020529.0-115925~A  & \ldots              &   242\footnotemark[1] & \ldots   &  4.2     &  3.9     \\
DENIS-P~J020529.0-115925~B  & \ldots              &   247\footnotemark[1] & \ldots   &  7.5     &  5.1     \\
DENIS-P~J035726.9-441730~A  & $<$-3               &   17.4                &   12.6   & \ldots   &  2.9     \\
DENIS-P~J035726.9-441730~B  & $<$-6               &   12.4                &   10.9   & \ldots   &  \ldots  \\
DENIS-P~J100428.3-114648~A  & $<$-3               &   26.1                &   48.4   & \ldots   &  \ldots  \\
DENIS-P~J100428.3-114648~B  & $<$-3               &   10.9                &   26.6   & \ldots   &  \ldots  \\
DENIS-P~J122821.6-241541~A  & $<$-5     & 258\footnotemark[1] & \ldots   &  12.4:\footnotemark[2]  &  4.5  \\
DENIS-P~J122821.6-241541~B  & $<$-5   & 255\footnotemark[1] & \ldots &   
6.6:\footnotemark[2]  &  2.2:\footnotemark[2] \\
DENIS-P~J144137.3-094559~A  & $<$-3       &   53.6     &   17.5   & 4.3:\footnotemark[2]   &  \ldots  \\
GJ~569~Ba                  & $<$-3  &   13.8                &   16.6   & \ldots   &  \ldots  \\
GJ~569~Bb                  & $<$-3  &   22.1                &   19.3   & \ldots   &  \ldots  \\
2MASSW~J0746425+200032~A    &  -18.8                 &   27.1                &   17.4   & \ldots   &  \ldots  \\
2MASSW~J0746425+200032~B    & -19.1                 &   37.4                &   17.6   & \ldots   &  \ldots  \\
2MASSW~J0920122+351742     & \ldots                &   363\footnotemark[1] & \ldots   & \ldots   &  4.3     \\
2MASSW~J1146344+223052~A    & $<$-6                 &   160\footnotemark[1] & \ldots   &  3.5     &  2.4     \\
2MASSW~J1146344+223052~B    & $<$-7   &   168\footnotemark[1] & \ldots   &  4.6     &  4.1:\footnotemark[2] \\
2MASSW~J1311391+803222~A    & $<$-3             &   36.0                &   18.9   & \ldots   &  \ldots  \\
2MASSW~J1311391+803222~B    & $<$-3             &   22.7                &   18.4   & \ldots   &  \ldots  \\
2MASSW~J1426316+155701~A    & $<$-3  &   37.5                &   11.8   & \ldots   &  \ldots  \\
2MASSW~J1426316+155701~B    & $<$-3                &   51.9                &   21.4   & \ldots   &  \ldots  \\
\hline
\end{tabular}

\thanks{Note.--- Equivalent width values are in given angstr\"oms and wavelegths are given in nanometers. 
1-$\sigma$ uncertainties typically are $\sim$15 \% of the equivalent width.\\}
\thanks{\footnotemark[1] Corresponds to the blend of the 7665 and 7699~\AA\, doublet. Integration 
limits 734.0-787.0~nm. \\}
\thanks{\footnotemark[2] Uncertain measurement due to blending or noise}
\end{table*}

\begin{table*}
\caption[]{Molecular band and pseudocontinuum slope data}
\label{atomic_lines}
\begin{tabular}{ lccccccc}
\hline
\hline
 Object                    & CrH1 & FeH1 & H$_2$O & TiO1 & TiO2 & VO1  & VO2    \\  
\hline 
\hline
DENIS-P~J020529.0-115925~A  & 1.64 & 1.43 & 1.20   & 1.06 & 0.98 & 0.84 & 1.42  \\
DENIS-P~J020529.0-115925~B  & 1.07 & 0.97 & 1.22   & 1.14 & 0.86 & 0.88 & 1.37   \\
DENIS-P~J035726.9-441730~A  & 1.03 & 0.85 & 1.33   & 1.32 & 1.80 & 1.51 & 1.23   \\
DENIS-P~J035726.9-441730~B  & 1.10 & 1.33 & 1.23   & 0.97 & 1.09 & 1.05 & 1.29   \\
DENIS-P~J100428.3-114648~A  & 1.02 & 1.04 & 1.21   & 1.50 & 1.63 & 1.28 & 1.04   \\
DENIS-P~J100428.3-114648~B  & 0.97 & 0.98 & 1.15   & 1.55 & 1.41 & 1.23 & 1.44   \\ 
DENIS-P~J122821.6-241541~A  & 2.19 & 1.91 & 1.17   & 0.74 & 1.03 & 0.55 & 1.30   \\
DENIS-P~J122821.6-241541~B  & 1.73 & 1.66 & 1.06   & 0.66 & 0.98 & 0.65 & 1.50   \\ 
DENIS-P~J144137.3-094559~A  & 1.24 & 1.23 & 1.14   & 1.27 & 1.42 & 1.13 & 1.15   \\  
GJ~569~Ba                   & 1.05 & 0.99 & 1.24   & 1.55 & 1.57 & 1.25 & 1.25   \\
GJ~569~Bb                   & 0.98 & 0.99 & 1.19   & 2.10 & 1.67 & 1.33 & 1.11   \\
2MASSW~J0746425+200032~A    & 1.20 & 1.19 & 1.24   & 1.25 & 1.44 & 1.12 & 1.22   \\
2MASSW~J0746425+200032~B    & 1.39 & 1.40 & 1.18   & 1.04 & 1.22 & 1.01 & 1.25   \\
2MASSW~J0920122+351742      & 1.63 & 1.41 & 1.21   & 0.40 & 0.83 & 0.68 & 1.31   \\
2MASSW~J1146344+223052~A    & 1.52 & 1.53 & 1.17   & 0.80 & 1.12 & 0.90 & 1.31   \\
2MASSW~J1146344+223052~B    & 1.37 & 1.38 & 1.11   & 0.84 & 1.04 & 0.90 & 1.26   \\
2MASSW~J1311391+803222~A    & 1.06 & 1.03 & 1.22   & 2.35 & 1.71 & 1.26 & 1.35   \\
2MASSW~J1311391+803222~B    & 1.03 & 0.97 & 1.20   & 2.37 & 1.68 & 1.39 & 1.27   \\
2MASSW~J1426316+155701~A    & 1.11 & 1.09 & 1.23   & 2.07 & 1.90 & 1.28 & 1.17   \\
2MASSW~J1426316+155701~B    & 1.38 & 1.49 & 1.14   & 0.95 & 1.26 & 1.16 & 1.08   \\ 
\hline
\end{tabular}

\thanks{Note.--- Integration limits for these spectral indices are given in 
M99.\\}
\end{table*}

\clearpage

\begin{table*}
\caption[]{Comparison of optical spectral types}
\label{atomic_lines}
\begin{tabular}{ lccc }
\hline
\hline
 Object                     & This work & M99  & K99,00    \\  
\hline 
\hline
DENIS-P~J020529.0-115925~A  & dL5      & dL5   & L7          \\
DENIS-P~J020529.0-115925~B  & dL6      &       &             \\
DENIS-P~J122821.6-241541~A  & dL4      & bdL4.5  & L5        \\
DENIS-P~J122821.6-241541~B  & dL4.5    &         &           \\ 
DENIS-P~J144137.3-094559~A  & dL1      & dL1   &             \\  
GJ~569~Ba                   & dM9      & dM8.5 &             \\
GJ~569~Bb                   & dM9      &       &             \\
2MASSW~J0746425+200032~A    & dL0      &       & dL0.5       \\
2MASSW~J0746425+200032~B    & dL1.5    &       &             \\
2MASSW~J0920122+351742~A    & dL5      &       &             \\
2MASSW~J1146344+223052~A    & dL2      &       & L3          \\
2MASSW~J1146344+223052~B    & dL2      &       &             \\
2MASSW~J1311391+803222~A    & dM7.5    &       &             \\
2MASSW~J1311391+803222~B    & dM8      &       &             \\
2MASSW~J1426316+155701~A    & dM8      &       &             \\
2MASSW~J1426316+155701~B    & dL1.5    &       &             \\ 
\hline
\end{tabular}

\thanks{References: K99: Kirkpatrick et al. 1999; K00: Kirkpatrick et al. 2000; 
M99: Mart\'\i n et al. 1999.\\}
\end{table*}

\clearpage


   \begin{table*}
	      \caption[]{Spectral types and photometric magnitudes for binaries 
with known parallaxes.} 
         \label{obs_log}
     \begin{tabular}{lcccccc}
            \hline
	    \hline

            Object         & Sp.T.            & m-M            & M$_{F814W}$      
& M$_{J}$  & M$_{Ks}$    & Refs.    \\
           \hline
            (abridged name)        &  (adopted)           &            &      
&    &     &     \\
           \hline
\hline
DENIS~J0205-1159~A &                     dL5                    & 1.48$\pm$0.06  & 17.10$\pm$0.12
 &                               &  12.3$\pm$0.1 &  Bo03,Ko99     \\ 
DENIS~J0205-1159~B\footnotemark[2] &     dL6                    &          & 18.18$\pm$0.12 
 &                                &  12.3$\pm$0.1               &  Bo05,Ko99        \\ 
DENIS~J1228-2415~A &                     dL4                  & 1.53$\pm$0.01  & 16.62$\pm$0.04  
& 13.56\footnotemark[1]$\pm$0.08  &  12.0$\pm$0.1  & Br04,Ko99,MBB99,V04   \\
DENIS~J1228-2415~B &                    dL4.5                    &             & 16.71$\pm$0.04  
& 13.76\footnotemark[1]$\pm$0.08  &  12.0$\pm$0.1  & Ko99,MBB99                       \\ 
GJ~569~Ba                  &             dM9                    & -0.04$\pm$0.02 & 12.54$\pm$0.16  
& 11.18$\pm$0.08                  & 10.06$\pm$0.09 & L01                         \\
GJ~569~Bb                  &             dM9                     &               & 13.24$\pm$0.16           
& 11.69$\pm$0.08                  & 10.47$\pm$0.09 &                             \\
2MASS~J0746+2000A    &                   dL0                     & 0.43$\pm$0.01  & 14.98$\pm$0.16  
& 11.76$\pm$0.08                  & 10.60$\pm$0.03 & Bo04                        \\
2MASS~J0746+2000B    &                   dL1.5                     &              & 15.98$\pm$0.21  
& 12.36$\pm$0.22                  & 11.12$\pm$0.05 & Bo04                        \\
2MASS~J1146+2230~A   &                   dL2                     & 2.83$\pm$0.05  & 15.34$\pm$0.17 
&                                 & 10.4$\pm$0.1              &  Ko99,Bo03,V04  \\
2MASS~J1146+2230~B   &                   dL2                     &                   & 16.09$\pm$0.22 
&                                 & 10.6$\pm$0.1                &  Bo03, Ko99              \\
            \hline
     \end{tabular}

\thanks{References: Br04=Brandner et al. 2004, Bo03=Bouy et al. 2003, 
Bo04=Bouy et al. 2004, Bo05=Bouy et al. 2005, G02=Geballe et al. 2002, K99=Kirkpatrick et al. 1999, 
Ko99=Koerner et al. 1999, L01=Lane et al. 2001, MBB99=Mart{\'{\i}}n, Brandner \& Basri 1999, 
V04=Vrba et al. 2004. \\}
\thanks{\footnotemark[1] M$_{J}$ deduced from F110M data from Mart{\'{\i}}n et al. 2000.\\}
\thanks{\footnotemark[2] DENIS-P~J0205-1159~B is itself likely double (Bouy et al. 2005). 
Our STIS spectrum and 
the $Ks$ magnitude of Koerner et al. 1999 include the sum of the two components.}
   \end{table*}


\clearpage

   \begin{figure*}
   \centering
   \includegraphics[width=\textwidth]{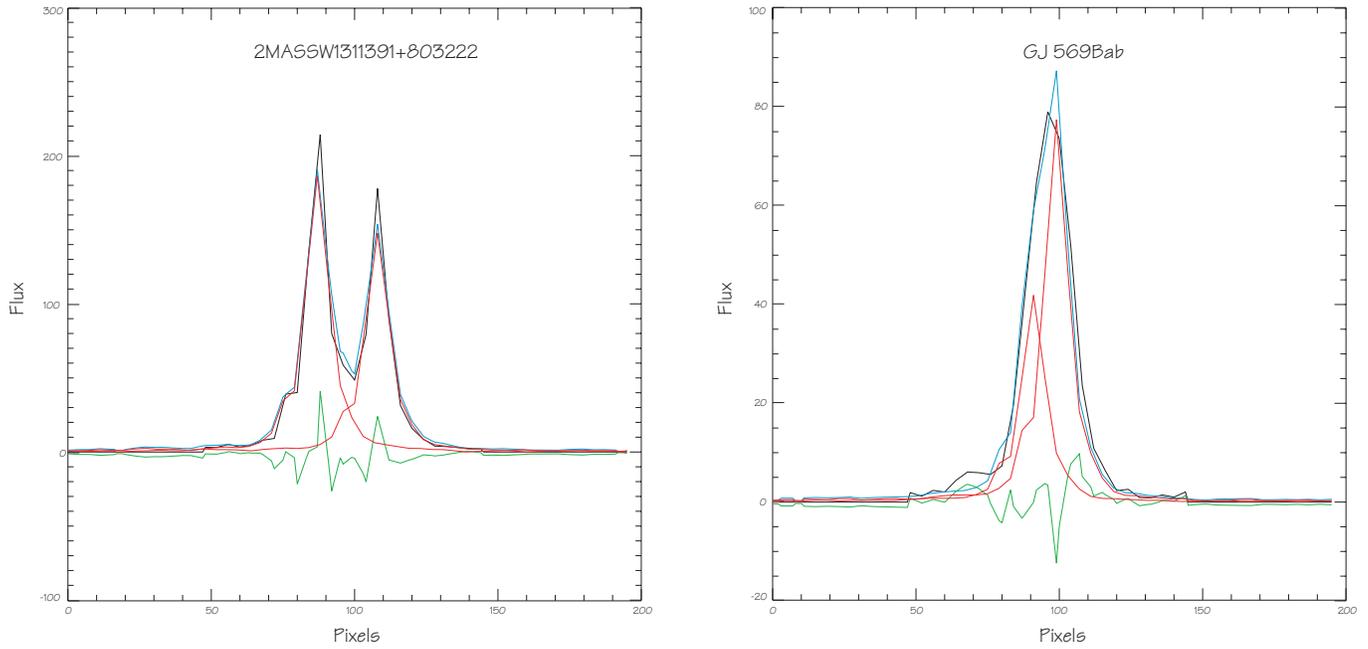}
   \caption{This figure shows the PSF fitting to the 2-dimensional 
STIS spectrum at 743.8~nm for  
two binary systems, namely 2MASSW~J1311391+803222 (separation = 0\farcs263) and 
 GJ~569~B (separation = 0\farcs090). The black lines show the observed profiles, 
the red lines indicate the best PSF fits for the components of these systems, 
the light blue line denotes the sum of the best PSF fits, and the green line shows 
the residuals obtained by substracting the sum of the PSF fits to the observed profile.
The integrated flux of the
residuals is about 3\% of the integrated flux of the total PSF.}
   \end{figure*}

\clearpage

   \begin{figure}
   \centering
   \includegraphics[width=\textwidth]{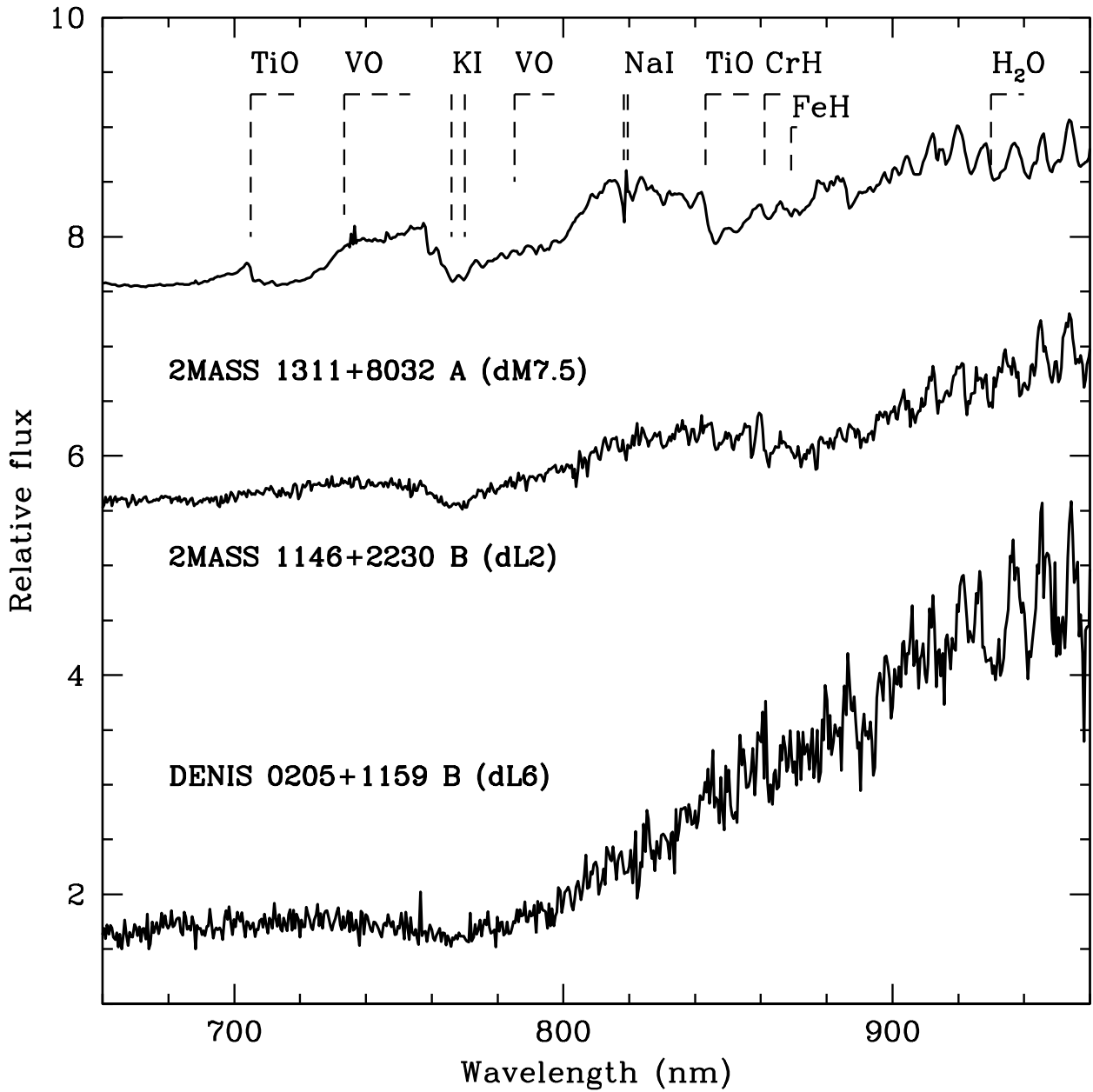}
   \caption{Final STIS spectra of three resolved ultracool dwarf binaries 
covering a representative range 
of spectral subclasses. The main spectral features identified  
in these spectra are labelled.}
   \end{figure}

\clearpage

   \begin{figure}
   \centering
   \includegraphics[width=\textwidth]{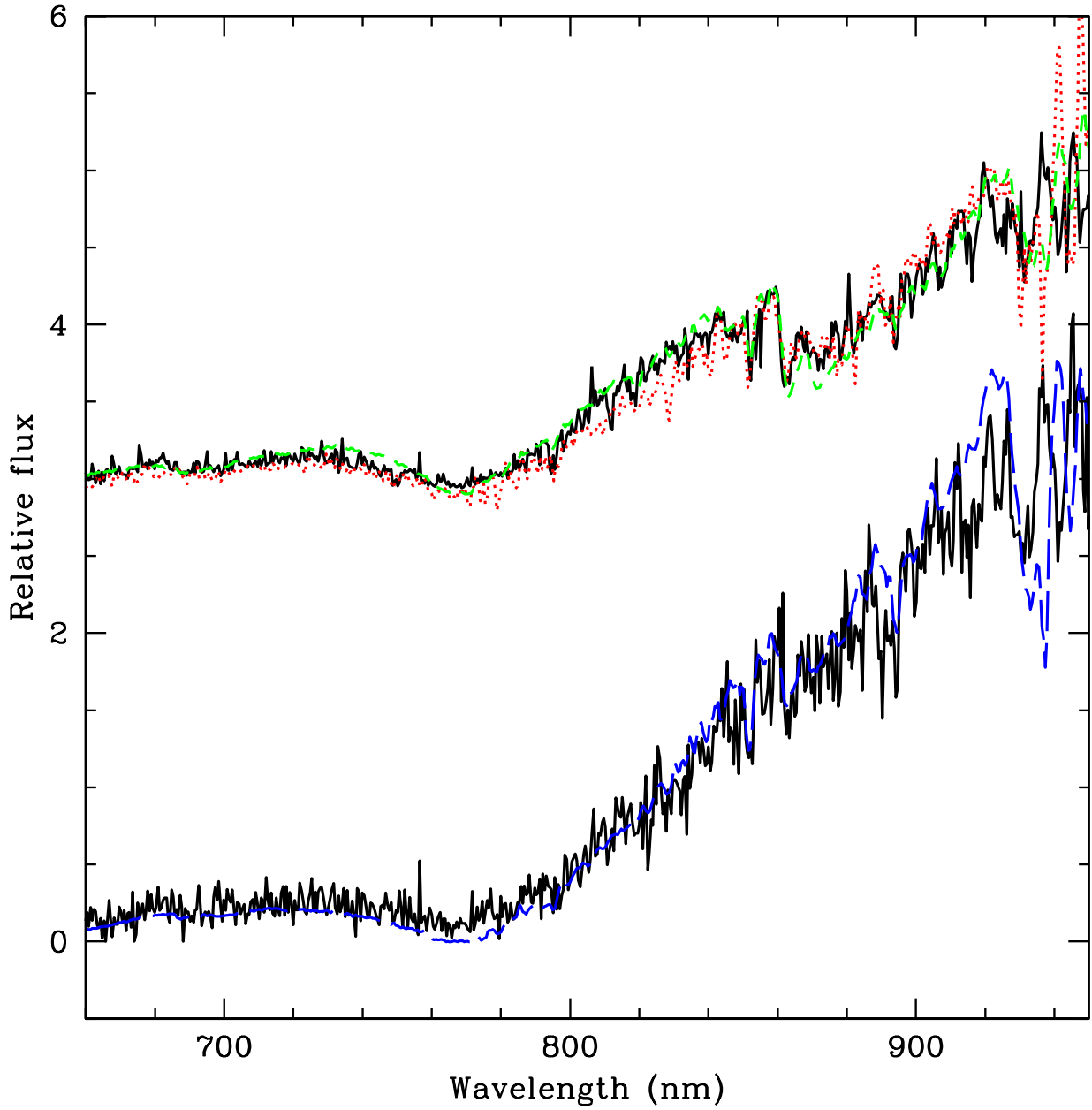}
   \caption{Final STIS spectra of the components of the binary 
DENIS~J0205-1159~A (top) and B (bottom). The following ground-based spectra of M99 
are shown for comparison: DENIS~J0205-1159 (dL5, red dotted line), 
DENIS~J1228-1547 (bdL4.5, green 
short dashed line), and DENIS~J0255-4700 (dL6, blue long-dashed line).}
   \end{figure}

\clearpage

   \begin{figure}
   \centering
   \includegraphics[width=\textwidth]{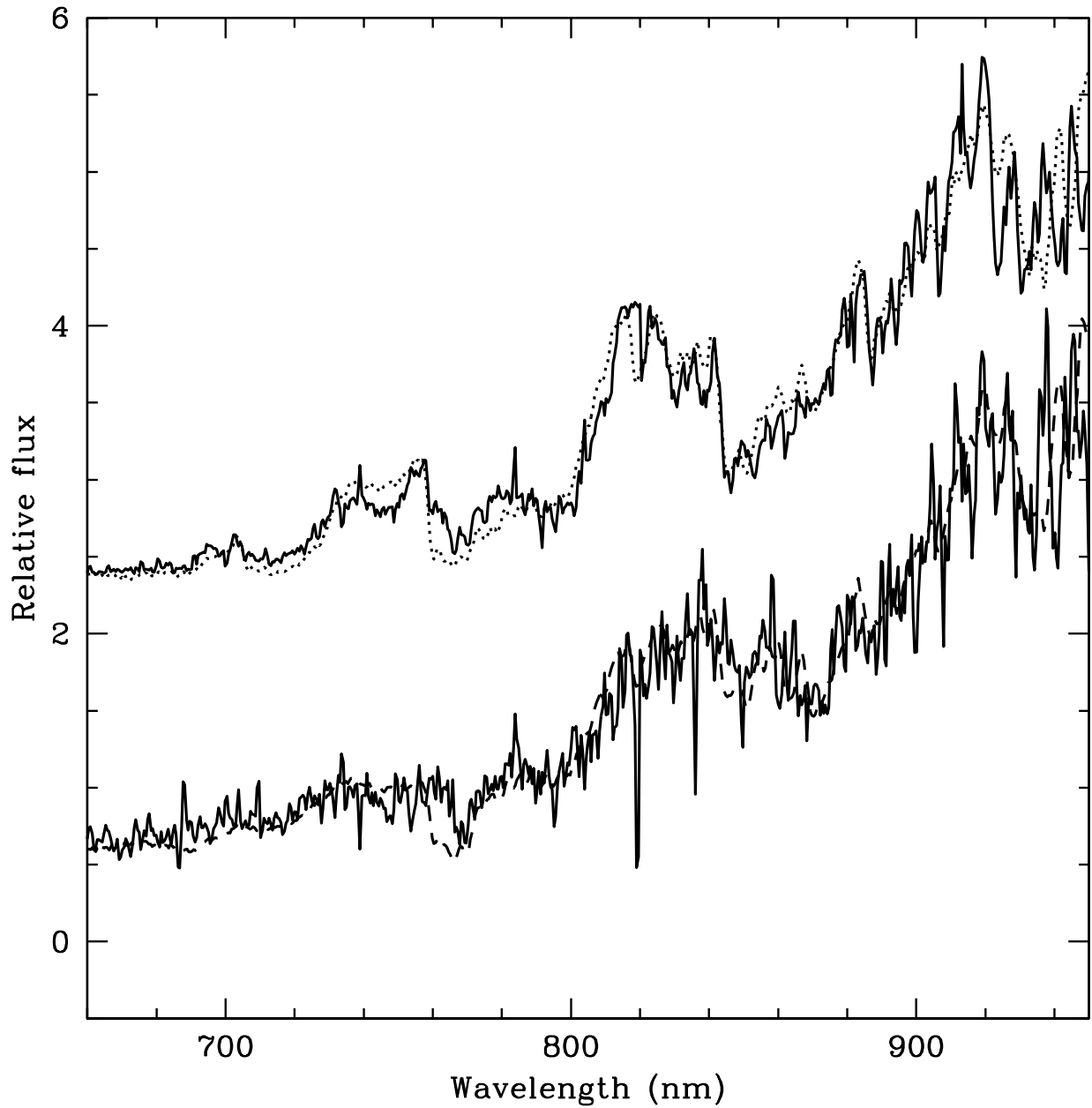}
   \caption{Final STIS spectra of the components of the binary 
DENIS~J0357-4417~A (top) and B (bottom). The following M99 spectra  
are shown for comparison: DENIS-P~1431-1953 (dM9, dotted) and DENIS~J1441-0945 
(dL1, dashed).}
   \end{figure}

\clearpage

  \begin{figure}
   \centering
   \includegraphics[width=\textwidth]{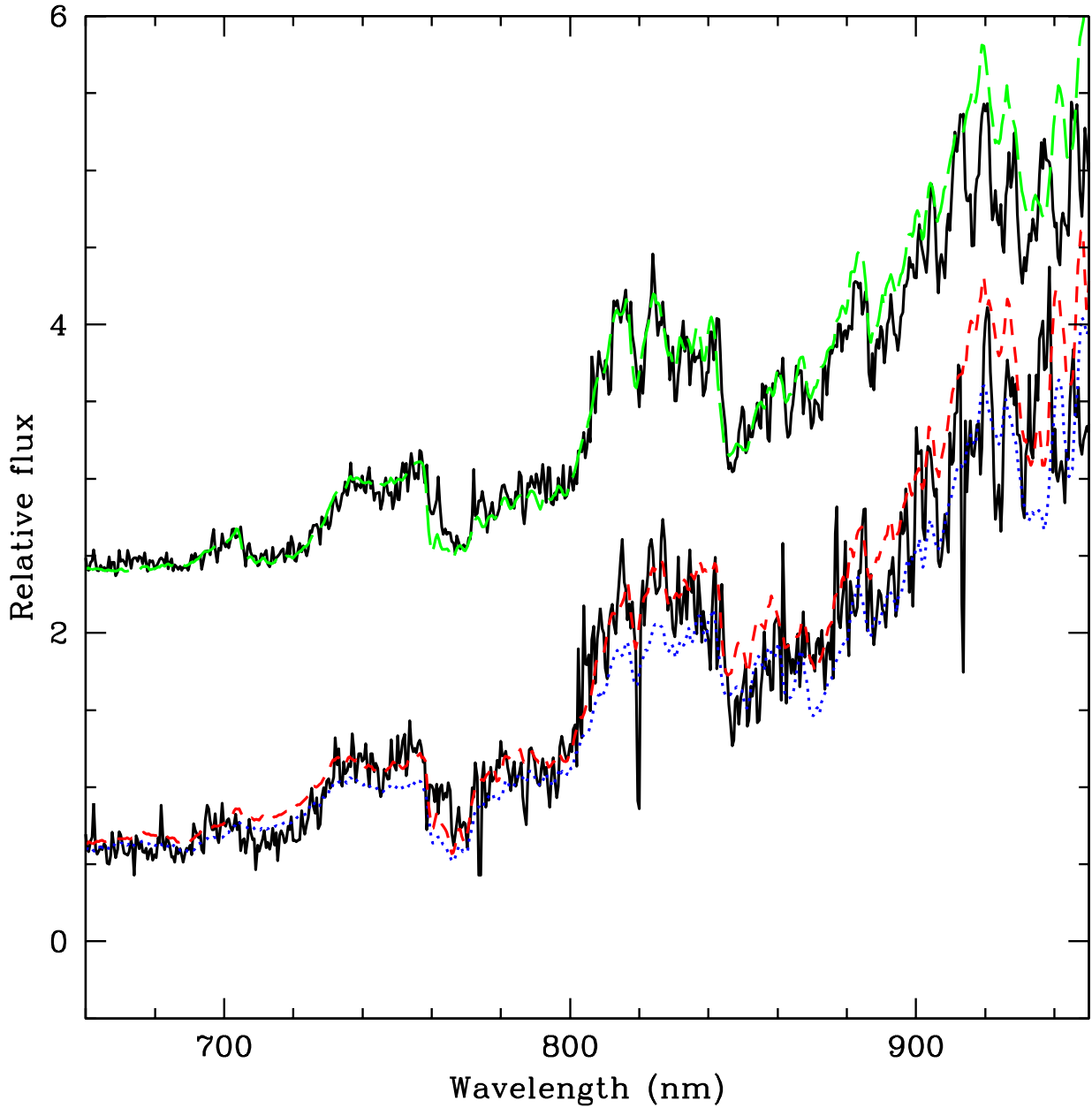}
   \caption{Final STIS spectra of the components of the binary 
DENIS~J1004-1146~A (top) and B (bottom). The following M99 spectra  
are shown for comparison: DENIS-P~1208+0149 (dM9.5, green long-dashed line),  
DENIS-P~0909-0658 (dL0, red short-dashed line), 
and DENIS~J1441-0945 (dL1, blue dotted line).}
   \end{figure}

\clearpage

   \begin{figure}
   \centering
   \includegraphics[width=\textwidth]{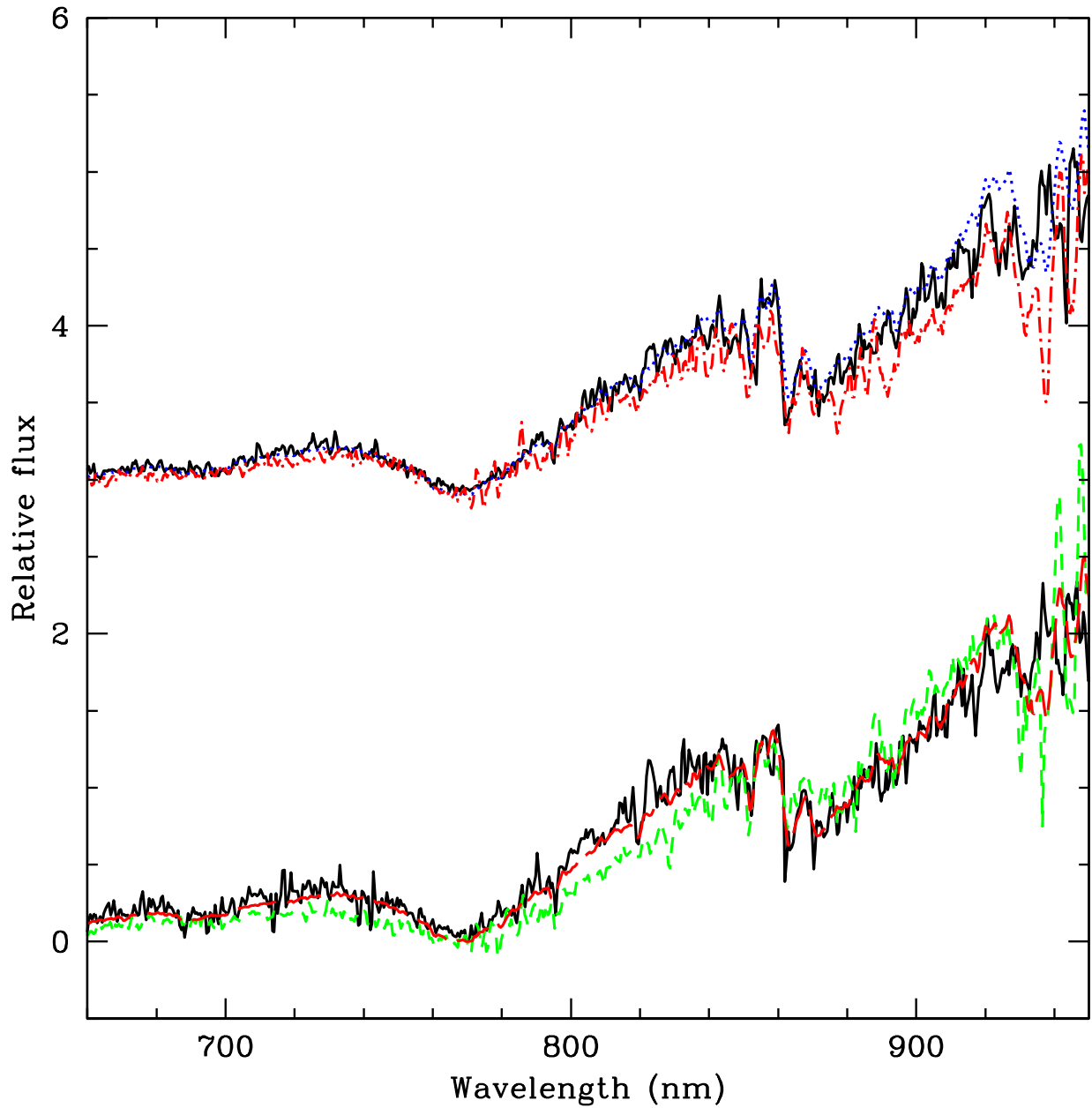}
   \caption{Final STIS spectra of the components of the binary 
DENIS~J1228-1547~A (top) and B (bottom). The following ground-based spectra of M99 
are shown for comparison: LHS~102~B (dL4, red dot-short dash line on top part of the plot), 
DENIS~J1228-1547 (bdL4.5, blue dotted line on top and red long-dashed line 
at the bottom), 
and DENIS~J0205-1159 (dL5, green short-dashed line).}
   \end{figure}

\clearpage

   \begin{figure*}
   \centering
   \includegraphics[width=\textwidth]{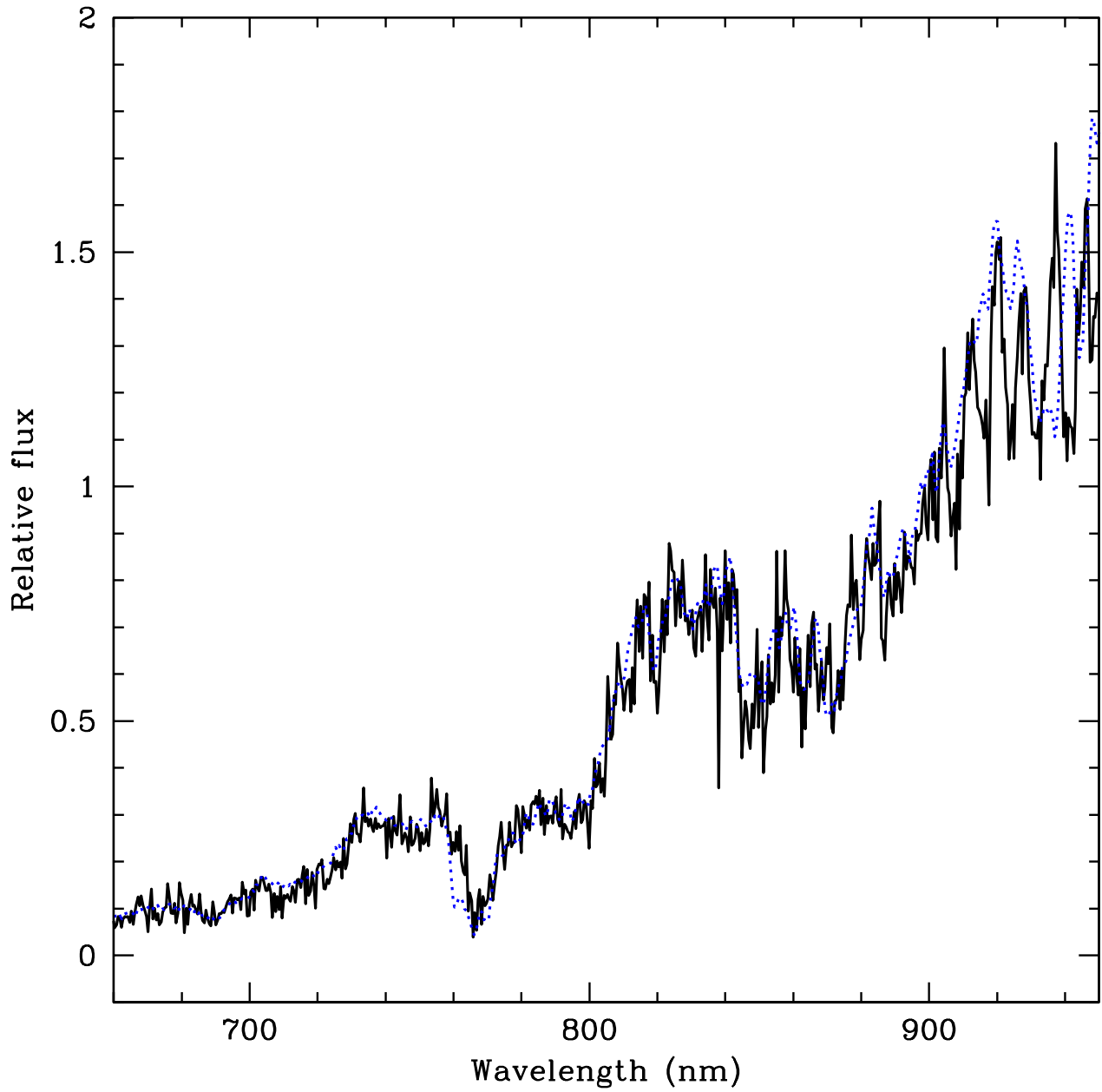}
   \caption{Final STIS spectrum of  
DENIS~J1441-0945~A compared with the M99 spectrum of DENIS~J1441-0945 
(dL1, blue dotted line).}
   \end{figure*}

\clearpage

   \begin{figure*}
   \centering
   \includegraphics[width=\textwidth]{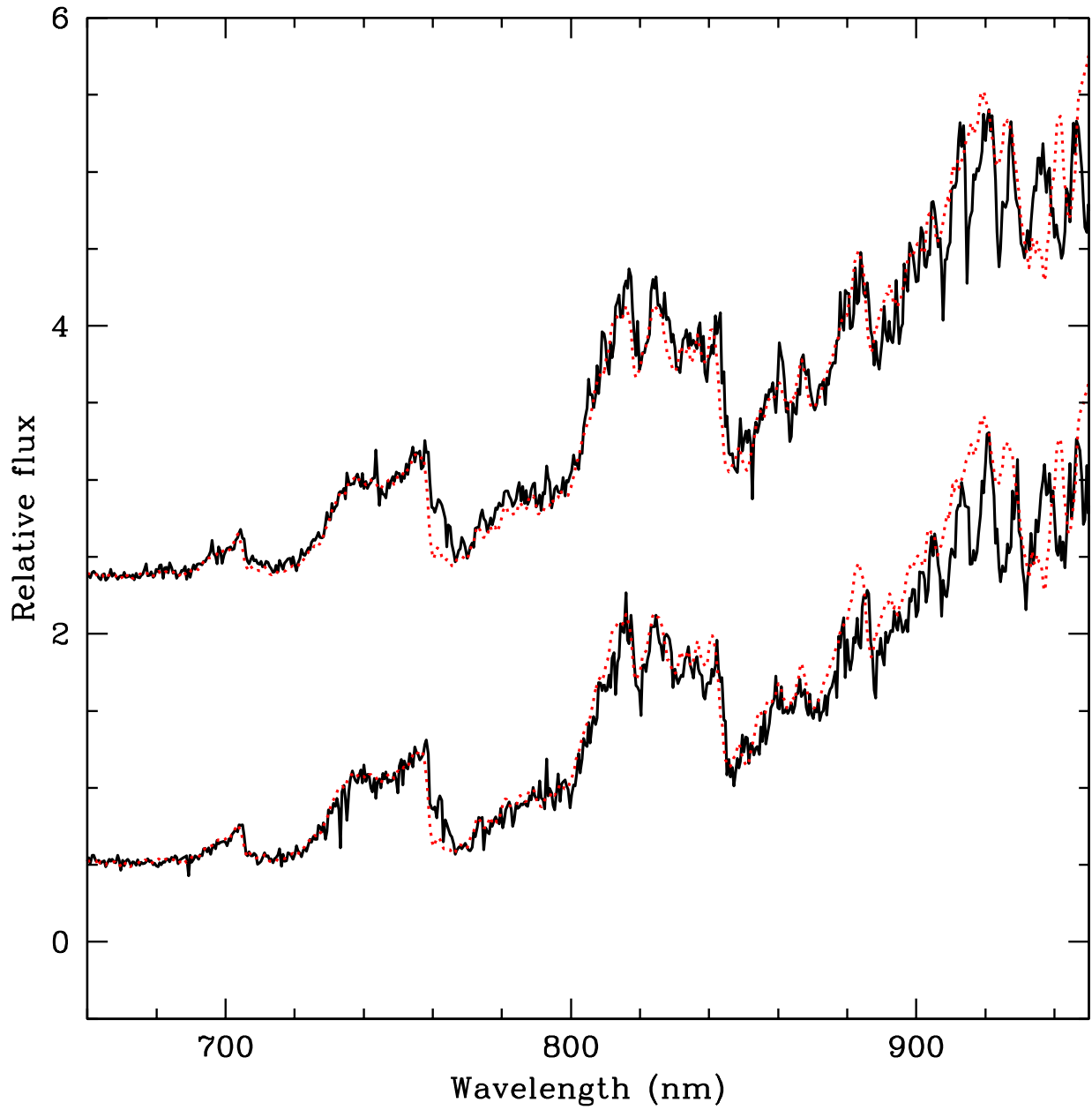}
   \caption{Final STIS spectra of the components of the binary 
Gl~569~Ba (top) and Bb (bottom), which are both matched  
with the M99 spectrum of DENIS-P~1431-1953 (dM9, red dotted line). }
   \end{figure*}

\clearpage

   \begin{figure*}
   \centering
   \includegraphics[width=\textwidth]{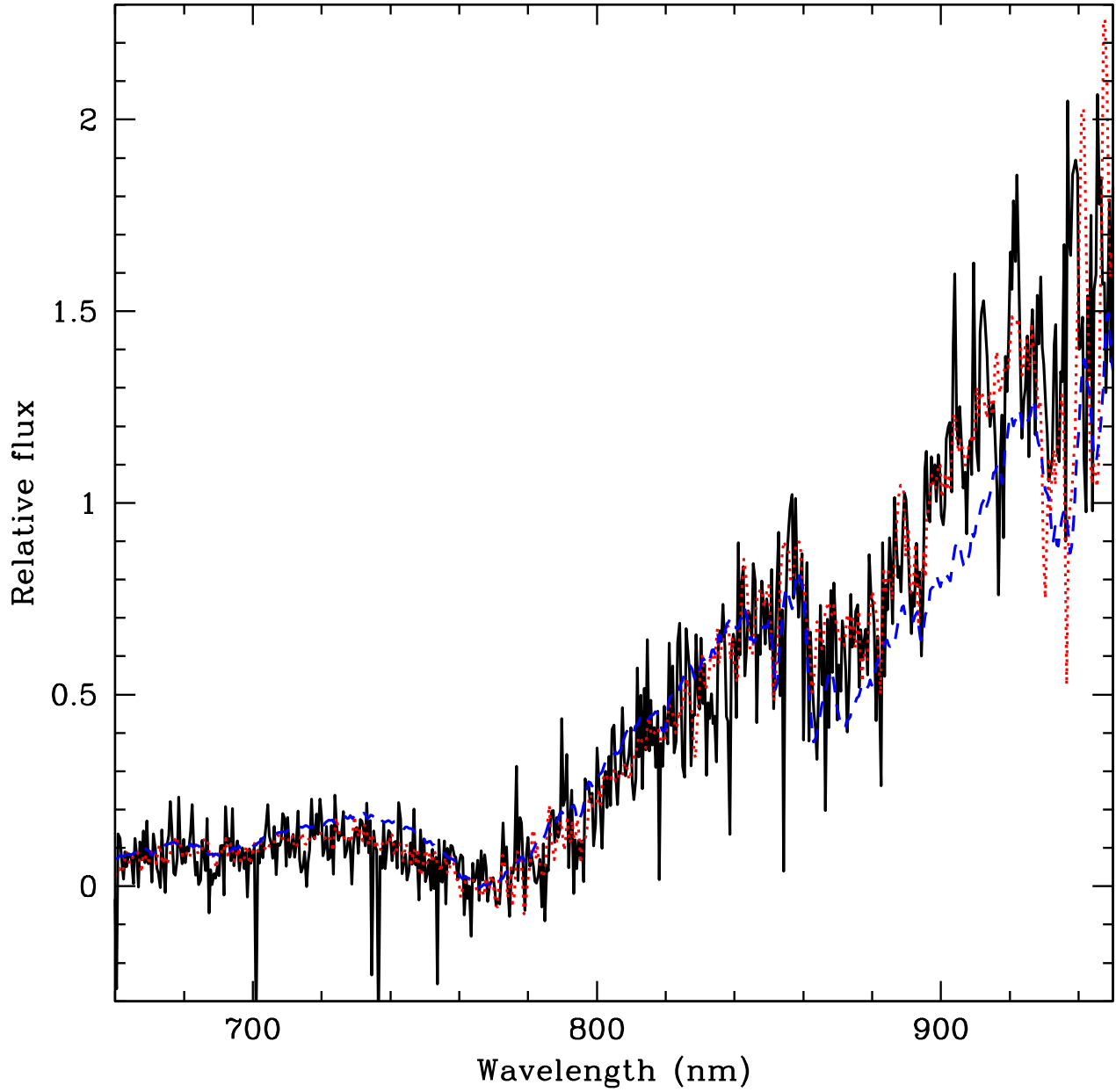}
   \caption{Final STIS spectrum of  
2MASS~J0920+3517~A compared with the M99 spectra of 
DENIS~J0205-1159 (dL5, red dotted line), and DENIS~J1228-1547 (bdL4.5, 
dashed blue line).}
   \end{figure*}

\clearpage

   \begin{figure*}
   \centering
   \includegraphics[width=\textwidth]{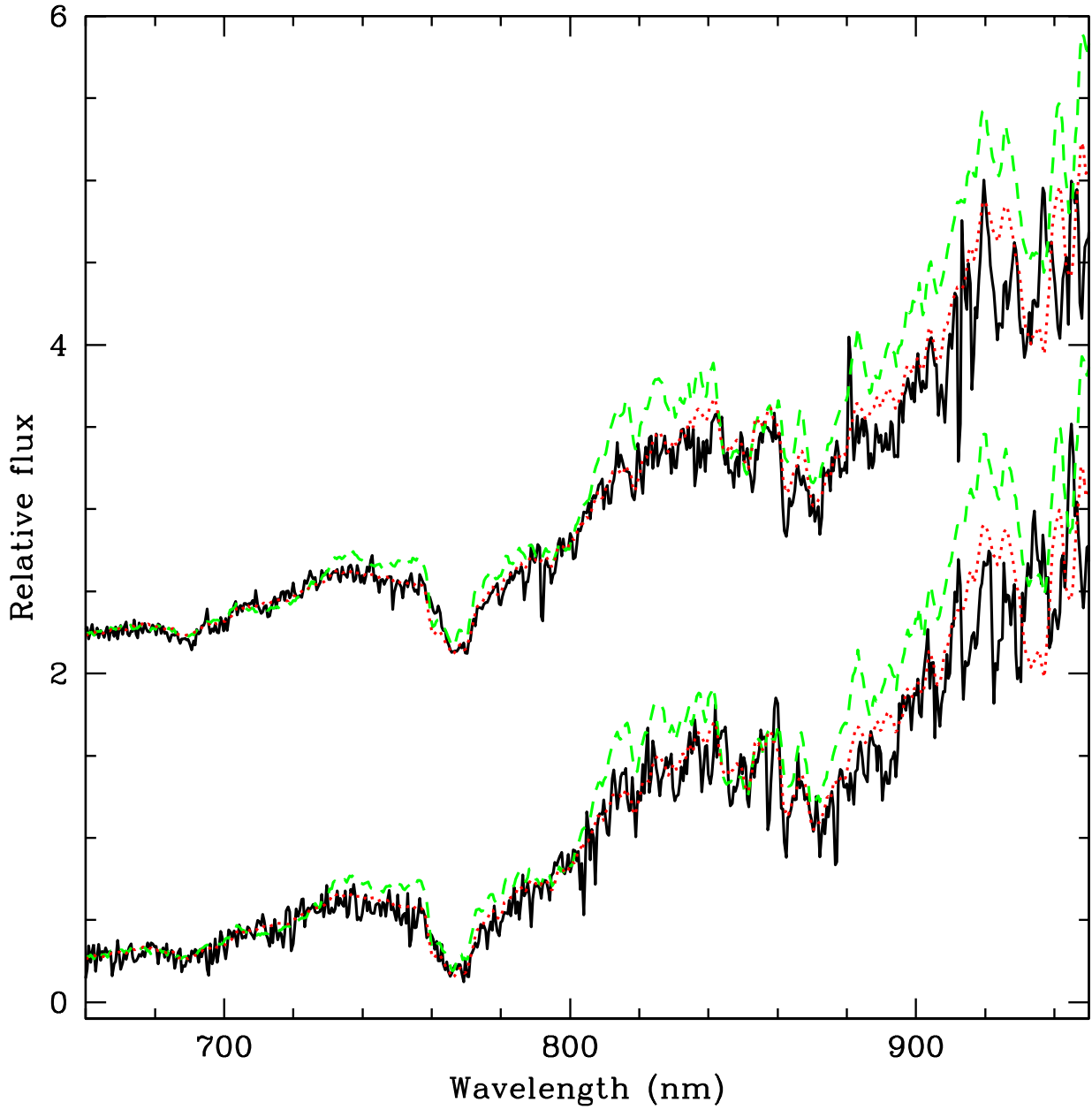}
   \caption{Final STIS spectrum of  
2MASS~J1146+2230~A (top) and B (bottom) compared with the M99 spectra of 
Kelu~1 (bdL2, red dotted line), and DENIS~J1441-0945 (dL1, 
dashed green line).}
   \end{figure*}

\clearpage

   \begin{figure}
   \centering
   \includegraphics[width=\textwidth]{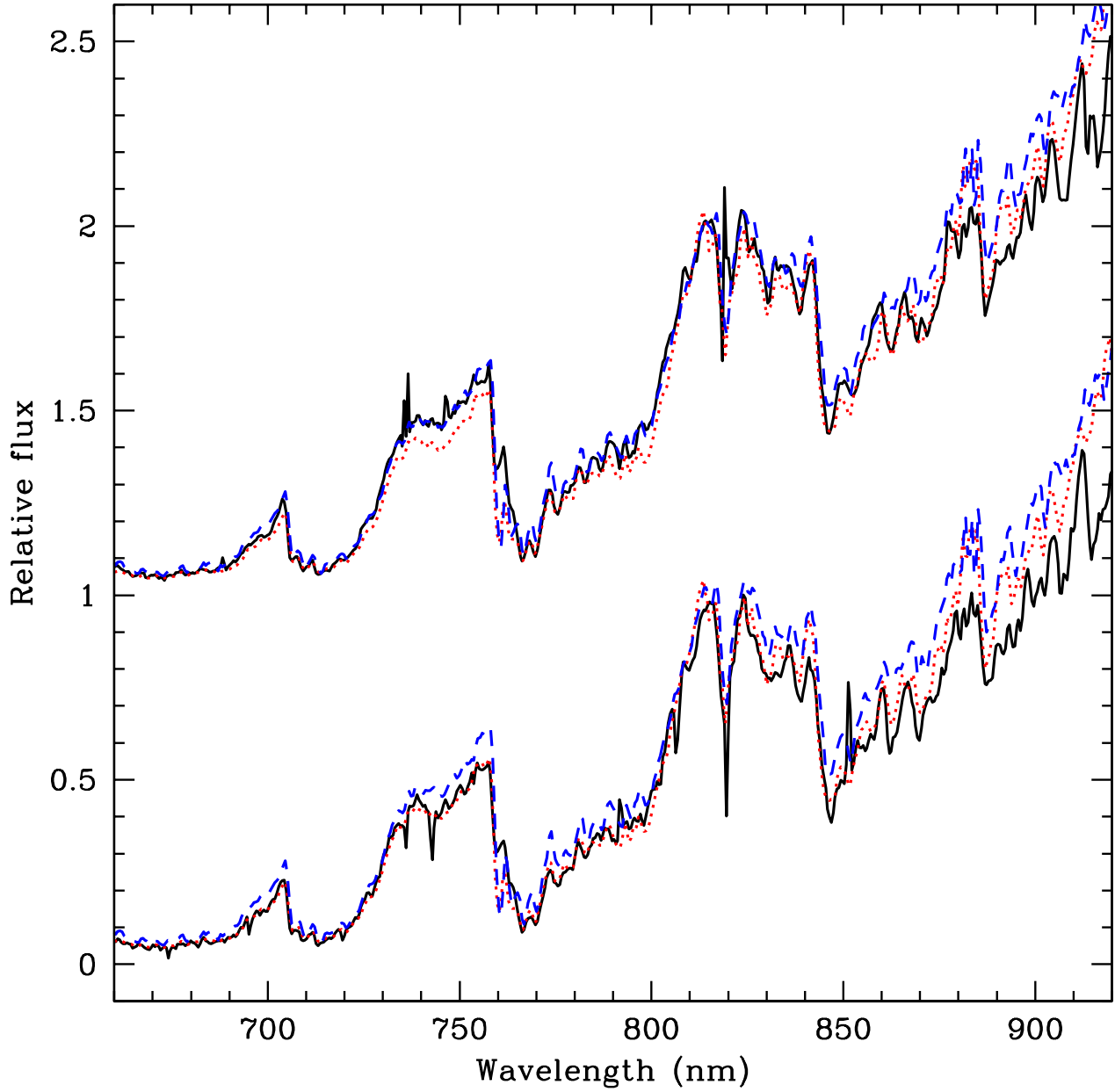}
   \caption{Final STIS spectrum of  
2MASS~J1311+8032~A (top) and B (bottom) compared with the M99 spectra of 
VB10 (dM8, red dotted line), and LHS2243 (dM7.5, dashed blue line).\label{sp2m1311}}
   \end{figure}

\clearpage

   \begin{figure*}
   \centering
   \includegraphics[width=\textwidth]{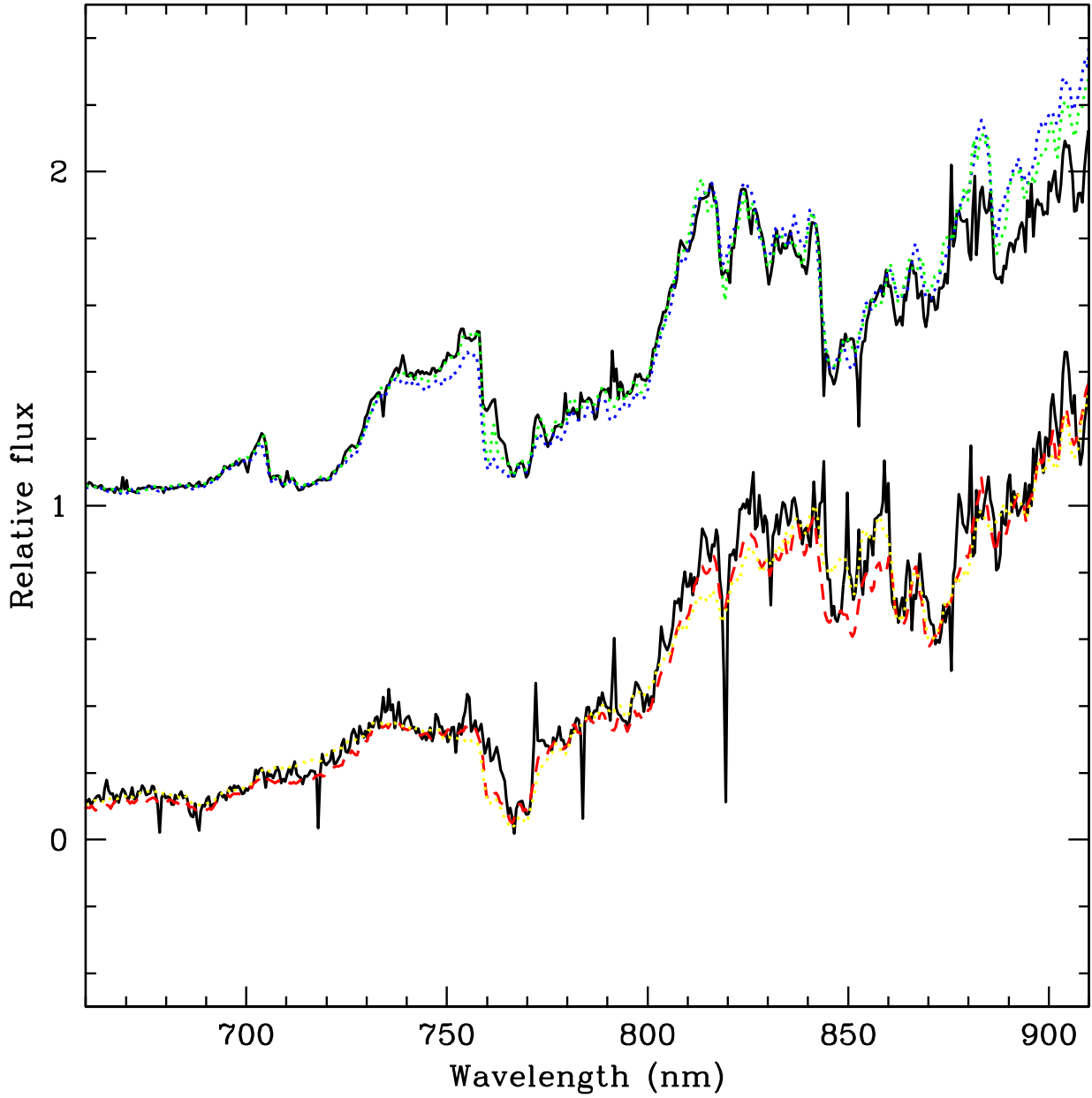}
   \caption{Final STIS spectrum of  
2MASS~J1426+1557~A (top) and B (bottom) compared with the M99 spectra of 
VB10 (dM8, green dotted line), DENIS-P~1431-1953 (dM9, blue dotted line), 
DENIS~J1441-0945 (dL1, dashed red line) and Kelu~1 (bdL2, yellow dotted line).\label{sp2m1426}}
   \end{figure*}

\clearpage

   \begin{figure*}
   \centering
   \includegraphics[width=\textwidth]{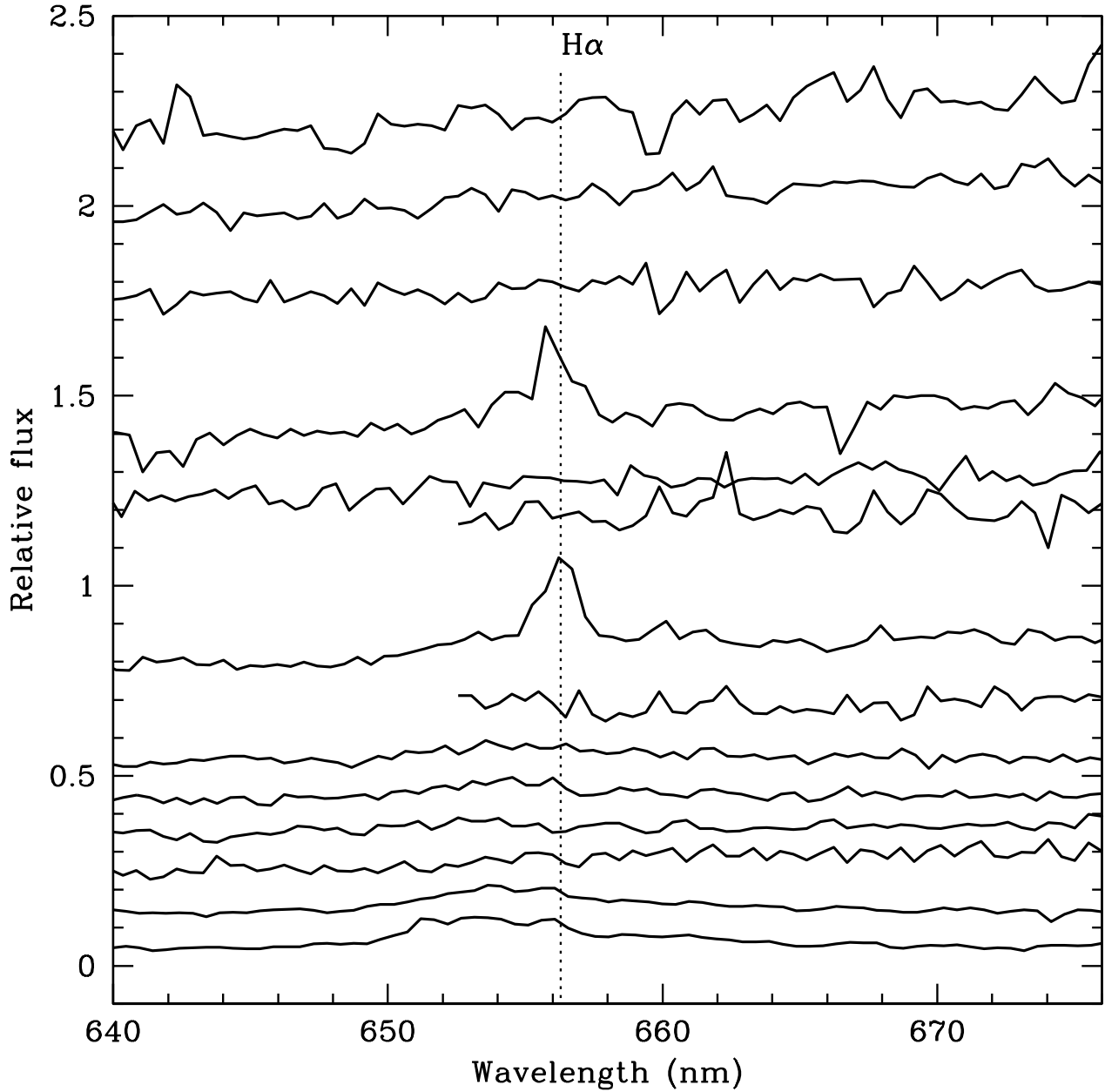}
   \caption{A zoom of the STIS spectral region around the H$_\alpha$ emission 
line. In order of increasingly late spectral subclass from bottom to top: 
2MASS~J1311+8032A and B, 2MASS~J1146+2230~A, DENIS-P~0357-4417~A, Gl~569~Ba and Bb, 
 DENIS~J1004-1146~A, 2MASS~J0746+2000~A, DENIS~J1004-1146~B, DENIS-P~1441-0945~A, 
2MASS~J0746+2000~B, 
2MASS~J1146+2230~B, DENIS~J1228-1547~A and B. \label{spHa}}
   \end{figure*}

\clearpage

  \begin{figure*}
   \centering
   \includegraphics[angle=-90,width=\textwidth]{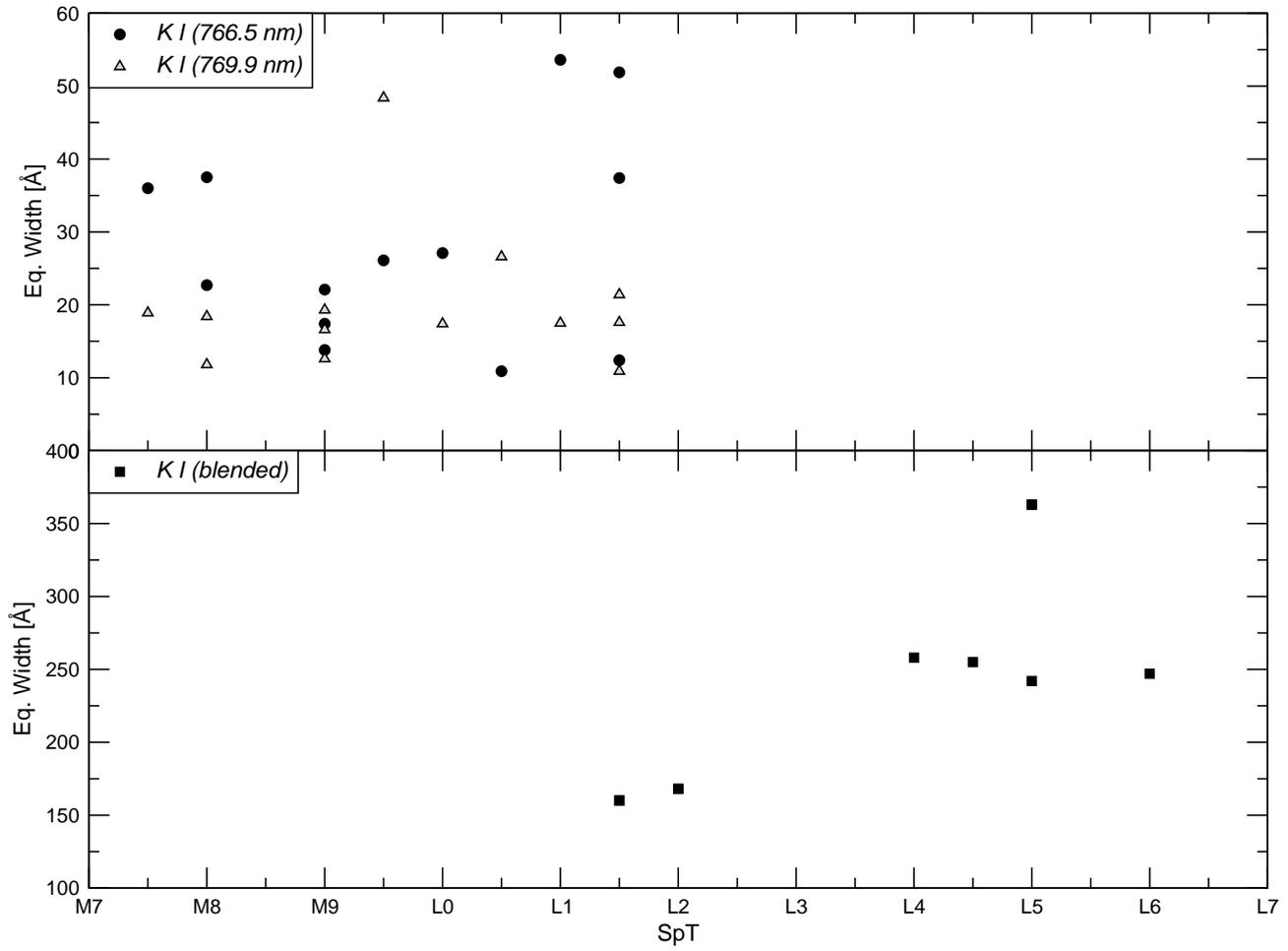}              
   \caption{Equivalent widths of the K~\textsc{i} resonance doublet at 766.5 and 769.9~nm
                                    versus spectral type for program ultracool dwarfs. }
   \end{figure*}
 
\clearpage

  \begin{figure*}
   \centering
   \includegraphics[angle=-90,width=\textwidth]{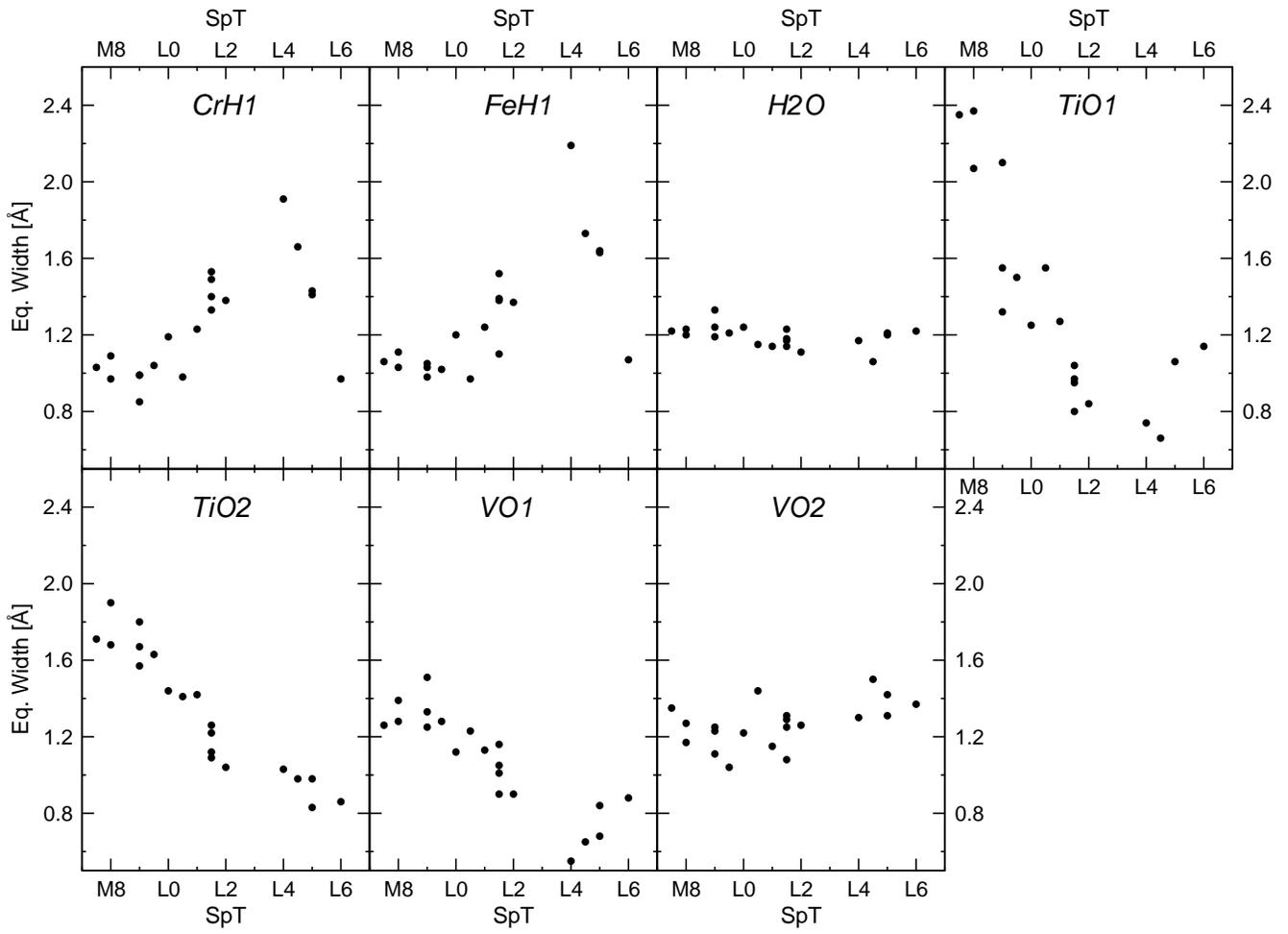}              
   \caption{Molecular absorption 
 indices  versus spectral type for program ultracool dwarfs. }
   \end{figure*}

\clearpage

  \begin{figure*}
   \centering
   \includegraphics[width=\textwidth]{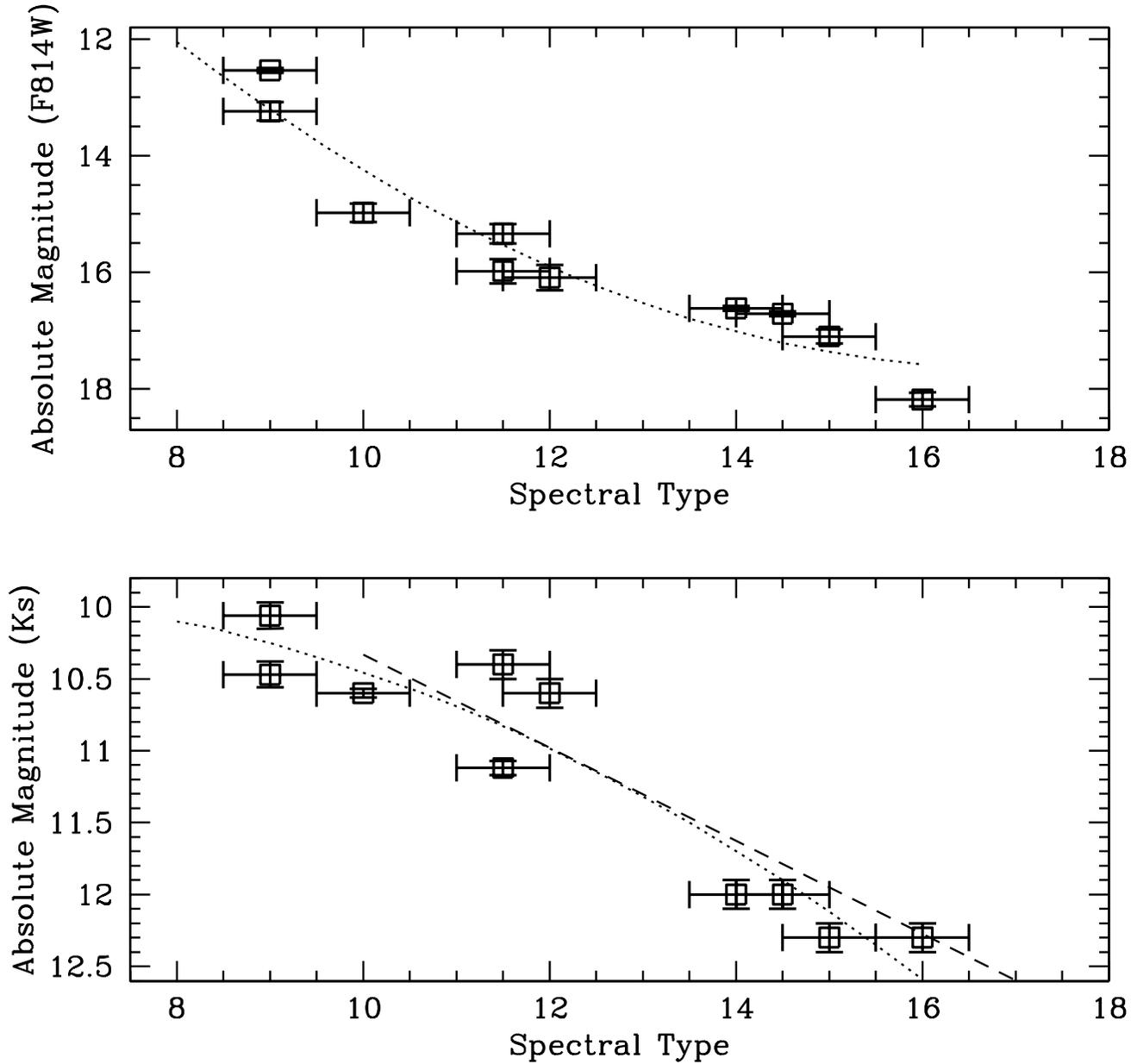}
   \caption{Absolute magnitude in the HST F814W filter (top panel) and K-band (bottom panel) 
versus spectral type for program ultracool 
 dwarfs. A second order polynomial fit to the data is shown with a dotted line. 
The dashed straight line represents the linear relatioship reported by Vrba et al. (2004).}
   \end{figure*}

\end{document}